\pdfoutput=1
\documentclass[11pt]{article}

\usepackage[letterpaper,margin=1in]{geometry}

\usepackage[T1]{fontenc}
\usepackage{textcomp}
\usepackage[mono=false]{libertine}
\usepackage{libertinust1math}
\usepackage[varqu,varl,scaled=0.95]{inconsolata}

\usepackage{microtype}

\usepackage{graphicx}
\usepackage{booktabs}
\usepackage{array}

\usepackage[font=small,labelfont=bf,labelsep=period]{caption}

\usepackage[numbers,sort&compress]{natbib}
\usepackage{xcolor}
\usepackage[colorlinks=true,
            linkcolor=black!50!blue,
            citecolor=black!50!blue,
            urlcolor=black!50!blue,
            breaklinks=true]{hyperref}

\providecommand{\BibTeX}{{\rm B\kern-.05em{\sc i\kern-.025em b}\kern-.08em T\kern-.1667em\lower.7ex\hbox{E}\kern-.125emX}}

\providecommand{\showDOI}[1]{\href{https://doi.org/#1}{\nolinkurl{doi.org/#1}}}
\providecommand{\showISBN}[1]{}
\providecommand{\showISSN}[1]{}
\providecommand{\showeprint}[2][]{\href{https://arxiv.org/abs/#2}{\nolinkurl{arXiv:#2}}}

\providecommand{\Description}[1]{}

\newenvironment{acks}{\par\vspace{0.5\baselineskip}\noindent\textbf{Acknowledgments.}\enspace\ignorespaces}{\par}

\usepackage{titlesec}
\titleformat{\section}{\normalfont\Large\bfseries}{\thesection}{0.75em}{}
\titleformat{\subsection}{\normalfont\large\bfseries}{\thesubsection}{0.6em}{}
\titlespacing*{\section}{0pt}{1.4\baselineskip}{0.5\baselineskip}
\titlespacing*{\subsection}{0pt}{1.0\baselineskip}{0.3\baselineskip}

\usepackage{placeins}

\begin{document}

\title{\bfseries Scalable and Personalized Oral Assessments\\Using Voice AI\thanks{This is the authors' version of a paper published in \emph{Communications of the ACM}. The definitive Version of Record is available at \url{https://doi.org/10.1145/3831714}.}}

\author{%
  Panos Ipeirotis\thanks{\href{mailto:panos@stern.nyu.edu}{\texttt{panos@stern.nyu.edu}}, ORCID 0000-0002-2966-7402.}\\
  Konstantinos Rizakos\thanks{\href{mailto:konstantinos.rizakos@nyu.edu}{\texttt{konstantinos.rizakos@nyu.edu}}. NiCE, Hoboken, NJ, USA; work done at New York University.}\\
  \normalsize New York University
}
\date{}

\maketitle

\begin{abstract}
\noindent
Written work no longer certifies that a student understands it: a polished
analysis now says little about who did the thinking. Oral examinations restore
that evidentiary link, but they have never scaled, because conducting and
grading them is expensive. We report on a system in which voice AI conducts a
personalized oral exam and a council of three large language models (LLMs)
grades the transcript, each model scoring independently and then revising after
reading the others. Across two undergraduate cohorts at NYU Stern (36 students
in Fall 2025, 37 in Spring 2026), a voice subscription covered all speaking
time and grading stayed under one dollar per exam. The deployments yield five
practical engineering lessons that should generalize wherever understanding
must be tested under questioning, from job interviews to professional
certification.

\end{abstract}

\section{The Problem}
\label{sec:problem}

Students in our AI/ML Product Management course at NYU Stern submitted thoughtful
project analyses and case-study discussions. The writing was often good:
clear arguments, reasonable tradeoffs, appropriate citations. Yet when we questioned
them live in class, many could not walk through a single decision from their own work.
The gap between what students submitted and what they could discuss live was too
consistent to be nerves.

The temptation is to call this a cheating problem~\cite{kasneci2023, denny2024},
but that framing misleads. AI-generated text is now difficult to distinguish from
human writing: in a controlled study, untrained readers identified GPT-3-authored text no better than a coin flip, and brief training raised accuracy to only about
55\%~\cite{clark2021}. Detection classifiers fail often enough that responsible
institutions have largely stopped using them~\cite{openai2023classifier,
vanderbilt2023turnitin}. The root issue is structural: written submissions once
served as evidence of thinking, and that evidentiary link has broken. Proctored,
closed-book exams restore something, but the open-ended, project-specific questions
that make a course like ours worth taking cannot survive a timed room with no
materials.

The same gap opens far beyond the classroom. A job interview, a professional
licensing board, even the legal question of whether two parties reached a
``meeting of the minds'': anywhere a polished artifact is taken as proof of
understanding that ought to be tested directly, the problem is the same, and so is
the remedy. Put the person in a position where they must reason aloud under
follow-up, and separate those who did the work from those who can only gesture at
it. We develop the idea inside a university course because that is where we
deployed it, but nothing about the method is specific to education.

Oral examinations sidestep the problem~\cite{fenton2025, hartmann2025}.
A student who can reason aloud through a follow-up question, defend a specific design
decision, and shift when challenged has demonstrated understanding in a way no artifact
can fake. Structured oral formats are also more reliable than their reputation suggests:
reliability concerns fall when examiners follow a shared rubric and pre-set
criteria~\cite{memon2010, nallaya2024}. Oral exams fell out of widespread use not
because they fail but because administering and grading them at scale cost too
much~\cite{bayley2024}. The same study reached a partial solution: 600 students recorded
video responses through a learning management system, but grading remained
human and the examination lacked the interactivity of human-driven exams. The scaling
problem stayed unsolved.

Voice AI and automated grading change that arithmetic. Both expensive parts of an oral
exam can now be automated: conducting it and grading the result. Contemporary
conversational AI platforms combine speech recognition, natural-sounding synthesis, and
turn-taking that tolerates thinking pauses~\cite{elevenlabs2024}.
Large language models (LLMs) can grade the resulting transcript reliably with the right
design~\cite{zheng2023}. Jabarian and Henkel~\cite{jabarian2025} found that AI-conducted
voice interviews in hiring can match or exceed human interviewers; Nitze~\cite{nitze2024}
explored LLM-driven oral simulations in text. Hung et al.'s Socratic Mind ran a separate 600-student pilot of an LLM-led Socratic oral-assessment assignment, taking spoken or typed
answers, with grading still manual~\cite{hung2024socratic}. Each piece already exists.

\begin{figure}[tbp]
\centering
\includegraphics[width=\textwidth]{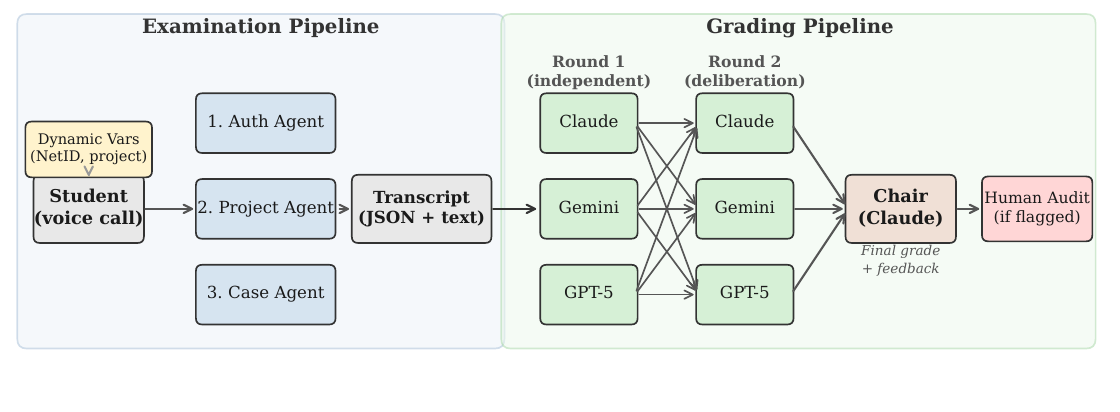}
\vspace{-0.5in}
\caption{End-to-end system architecture. \textbf{Left:} the examination pipeline runs three agent phases, with per-student context injected via dynamic variables. \textbf{Right:} the grading pipeline scores each transcript across two multi-model rounds, then a chair model synthesizes the final grade; high-disagreement cases route to human audit.}
\Description{Two-pipeline architecture. Left: a student voice-calls three sequential agents (Auth, Project, Case) under an orchestrator with dynamic variables, producing a transcript. Right: the transcript flows through Round 1 (Claude, Gemini, GPT-5 score independently), Round 2 (deliberation after peer scores), and a Chair (Claude) that synthesizes the final grade; flagged cases go to human audit.}
\label{fig:architecture}
\end{figure}

What no one had shown is how to assemble them into a live voice examination of record,
conducted and graded end to end, that holds up under real exam conditions, and what
breaks when you try. We built \emph{Viva by NYU Stern},\footnote{Available at
\url{https://viva.stern.nyu.edu/}. The name nods to ``viva voce,'' the Latin ``living
voice'' phrase for oral examinations.} which conducts a personalized oral exam and then
grades the transcript with a council of three LLMs that score independently, read each
other's assessments, and revise (Figure~\ref{fig:architecture}).

\paragraph{Scope and contribution}
This is a field report, not a controlled study. Two consecutive cohorts, one course,
one institution: 36 students in Fall 2025, 37 in Spring 2026. The numbers we
report (cost per student, how closely the grading models agree, rates of student stress
and preference) are observations from those deployments, not population estimates.
Whether the lessons that follow transfer beyond this one deployment is an empirical
question we cannot settle from two cohorts. The replicated-vs-cohort-specific split
appears in Section~\ref{app:replicated}.\footnote{The anonymized cohort data and the analysis scripts that reproduce the reported statistics are archived at Zenodo: \url{https://doi.org/10.5281/zenodo.21522314}; a sanitized snapshot of the orchestration and grading code accompanies the paper.}

What we offer are five practical engineering lessons that surfaced in our deployments and
are likely to matter in many conversational systems that hand behavior to a language
model. We call them \emph{patterns} in the design-pattern sense a programmer already
knows: each pairs a recurring failure with a reusable fix. The five:
(1)~\emph{Decompose into single-purpose modules, not a single
do-everything agent}; (2)~\emph{Constrain LLM behavior with code or configuration, not prompts};
(3)~\emph{Never delegate randomization to an LLM}; (4)~\emph{Use a multi-model council
with a deliberation round, and choose models that disagree}; and (5)~\emph{Choose voice
characteristics with the same care as question design}. The first pattern was an early
architectural choice; the others surfaced from deployment failures. These patterns
organize the rest of this paper. Education is where we found them.

\section{System Design}
\label{sec:system-design}

\subsection{Platform Requirements and Architecture}
\label{sec:platform}

We could have run the whole examination through a single agent. We split it across three instead: isolating each phase prevents conversational drift, keeps a failure in Phase~2 from contaminating Phase~3, and simplifies debugging. That design choice shaped what the platform had to provide.

A voice examination system needs four capabilities: (1)~accurate speech-to-text and text-to-speech with natural prosody, (2)~turn-taking that tolerates thinking pauses, (3)~multi-agent workflows to decompose the exam into phases, and (4)~transcript export for grading. These can be assembled from components or bought as an integrated platform. We used ElevenLabs Conversational AI~\cite{elevenlabs2024}, but the first, second, and fourth capabilities are commodity across contemporary voice-AI platforms (Retell, Vapi, Ultravox, LiveKit Agents, OpenAI's Realtime API); only multi-agent routing and per-session dynamic variables would require application-layer reimplementation elsewhere (Appendix~\ref{app:portability}).

Figure~\ref{fig:architecture} shows the flow, and the full prompt scaffold is in Appendix~\ref{app:voice-agent-prompt}. Each of the three agents owns one phase of the examination:

\begin{enumerate}
    \item \textbf{Authentication Agent:} In Fall 2025 this agent opened the session, captured the student's name and NetID through spoken dialogue, and validated them against the course roster before handing off to Phase~2. In Spring 2026 we made it largely redundant by issuing each student a personalized examination URL ahead of time: the link establishes identity, so the phase becomes an extra security check that a brief greeting could replace.
    \item \textbf{Project Discussion Agent:} Receives the student's capstone project context via per-session dynamic variables and asks structured questions about their specific work: goals, data choices, modeling decisions, evaluation approaches, and failure modes. This phase targets the gap between LLM-generated written work and actual understanding: students cannot improvise consistent answers about decisions in work they did not personally complete.
    \item \textbf{Case Discussion Agent:} Discusses one case study from a library of real-world AI/ML product cases covered in the semester, asking questions that span course topics. In Fall 2025 the agent itself selected the case from a prompt-supplied library; in Spring 2026 the case is drawn in code from a per-session random seed and passed to the agent as a dynamic variable (Section~\ref{sec:failure-modes}). The agent generates its questions dynamically during the examination.
\end{enumerate}

Only per-session dynamic variables flow across handoffs, not prior-phase transcript text; the downstream grading council (Section~\ref{sec:grading}) still sees the full concatenated transcript. The failure modes in Section~\ref{sec:failure-modes} originate in the underlying LLM, not the voice platform, so the patterns that address them surface on any stack that delegates conversation management to a language model (voice selection travels less easily; Appendix~\ref{app:portability}).

\subsection{Grading: A Council of LLMs}
\label{sec:grading}

We implemented a ``council of LLMs'': three models from different families independently score the same transcript, then revise after seeing each other's reasoning. We used Claude, Gemini, and GPT-5, one from each of the three major frontier-model families available at deployment rather than a tuned panel size.\footnote{Exact model identifiers, from our deployment configuration (Fall 2025) and the stored grade records (Spring 2026): Fall 2025 used \texttt{claude-opus-4-6}, \texttt{gemini-2.5-pro}, and \texttt{gpt-5.4}; Spring 2026 used \texttt{claude-opus-4-6}, \texttt{gemini-3.1-pro-preview}, and \texttt{gpt-5.4} (Gemini was the only model upgraded between cohorts). We record the exact versions because model version affects reproducibility; the chair-model tier, which differed between cohorts, is discussed in Appendix~\ref{app:cost}.} This ``LLM-as-judge'' approach extends a long history of automated scoring~\cite{shermis2013} to oral examination transcripts: language models evaluate open-ended work in place of human raters~\cite{zheng2023, verga2024} and can debate their judgments in multi-agent exchanges~\cite{chateval2023}. The process has three steps:

\begin{enumerate}
    \item \textbf{Round 1 (Independent):} Each model receives the full examination transcript and the grading rubric, then independently scores it on the rubric's dimensions (in our case, five dimensions, each scored 0--4), with verbatim transcript evidence for each score: \emph{Problem Framing} (business problem $\to$ ML spec), \emph{Metrics \& Economics} (evaluation metrics, trade-offs, counter-metrics), \emph{Risk \& Ethics} (failure modes; fairness, accountability, transparency, privacy, security), \emph{Experimentation} (A/B design, hypotheses, controls), and \emph{Communication} (structured answers under follow-up pushback). The three models run in parallel with no shared context.
    \item \textbf{Round 2 (Deliberation):} Each model receives the other two models' \emph{complete} Round~1 output (their per-dimension scores, written justifications, and the verbatim transcript excerpts each one cited) rather than a paraphrased digest, and is asked to revise its own evaluation in light of that reasoning. We label peers neutrally (``Peer 1,'' ``Peer 2'') and withhold model identities, so a model cannot defer to a brand rather than an argument.\footnote{We concatenate the two peer assessments in a fixed order for every exam; we did not randomize presentation order, so a small position effect is possible, though with both peers always present there is little room for order to matter.} Models may change scores or reaffirm them, but must justify either decision with transcript evidence. On one Spring 2026 exam (a DocuSign ``Contract Risk Analyzer'' project), a grader that had scored Problem Framing higher revised it down to 3 once a peer noted the student had never worked through the project's second-order effects, arguing the change from what the transcript showed rather than from the peer's number.
    \item \textbf{Chair Synthesis:} A designated chair model (Claude) receives all Round~1 and Round~2 outputs and produces the final grade: one score per dimension, a total score (0--20), and a structured feedback report with strengths, weaknesses, and verbatim evidence.
\end{enumerate}

Why not use a single model? Because initial independent assessments showed poor agreement; the deliberation phase was necessary. A known risk is \emph{anchoring bias}: when models see each other's Round~1 scores, they may converge toward a shared (possibly wrong) answer rather than genuinely revising. We cannot rule this out, but the deliberation transcripts show substantive engagement: models cite specific transcript evidence to argue for score changes rather than splitting differences. The convergence reads as reasoning, not anchoring. More on this in Section~\ref{sec:results}. The full grading prompt and a worked deliberation example are in Appendix~\ref{app:grading-prompt}.

\section{Deployment and Results}
\label{sec:results}

We deployed the system as the final examination for an undergraduate AI/ML Product Management course at NYU Stern across two consecutive cohorts, Fall 2025 (36 students) and Spring 2026 (37 students), with fixes shipped between them (Section~\ref{sec:failure-modes}). We use the semester labels throughout to line up with the dated tables and figures.

The system eliminated scheduling as a constraint; students chose any time during a nine-day window in each cohort (Table~\ref{tab:exam-stats} lists the opening dates). Students recorded themselves (webcam and audio) during the examination as a course requirement, both to deter outsourcing or real-time assistance and as a backup if the conversational pipeline failed mid-exam. This exam-integrity obligation is separate from research consent: under IRB protocol NYU IRB-FY2023-7595, de-identified transcripts and survey responses enter the analysis only for consenting students; a student who declined would still have taken the exam for course assessment, with their data excluded from the study. No student in either cohort declined.

\subsection{Cost}
\label{sec:cost}

\begin{table}[t]
\footnotesize
\caption{Examination Statistics, both cohorts. ``Messages'' counts both agent prompts and student replies.}
\label{tab:exam-stats}
\centering
\begin{tabular}{lrr}
\toprule
\textbf{Metric} & \textbf{Fall 2025} & \textbf{Spring 2026} \\
\midrule
Students examined & 36 & 37 \\
Exam window opens & Dec 12 & Mar 15 \\
Average duration & 22 min & 30 min \\
Duration range & 9--41 min & 19--43 min \\
Average number of messages & 58 & 84 \\
\bottomrule
\end{tabular}
\end{table}

Two regimes matter. \emph{Within} our \$99/month ElevenLabs subscription, which covered all voice minutes in both cohorts, the per-exam marginal cost reduces to grading-LLM API calls alone: ${\sim}\$0.29$ per Fall 2025 exam (billed) and ${\sim}\$0.96$ per Spring 2026 exam (estimated; Appendix~\ref{app:cost}), or \$0.01--\$0.03 per student-minute. The Spring jump is a configuration choice, not a scaling signal: a Claude Opus 4.6 chair accounted for roughly two-thirds of grading spend; reverting to a cheaper chair tier would halve the total. \emph{Above} an equivalent subscription credit pool, voice bills at the ElevenLabs additional-call overage rate of \$0.08 per agent-minute, so the planning figure is roughly \$0.11 per student-minute, or about \$3.40 for a 30-minute graded exam (before retry/setup overhead). For perspective, compare against human graders. On grading spend alone, both cohorts ran one to two orders of magnitude below a faculty/TA comparator of roughly \$750 (36 students $\times$ 25 minutes $\times$ two graders at \$25/hour): \$10.30 in Fall 2025 (${\sim}73\times$ cheaper) and \$35.48 in Spring 2026 (${\sim}21\times$ cheaper). The gap narrows once voice is fully priced but stays wide: at 1{,}000 students, the above-subscription rate adds roughly \$3{,}400 in inference against roughly \$25{,}000 for two human graders, about $7\times$ cheaper. Per-administration scaling is not gated by token or voice spend; the full decomposition, including the cohort-total reconciliation, is in Appendix~\ref{app:cost}.

These per-exam figures are the \emph{marginal} cost. The larger cost is a one-time \emph{adaptation}: writing the rubric, curating the case library, and authoring the per-phase questions. The orchestration and grading platform already exists (we now run it as a hosted service), so this is exam-authoring effort rather than a systems-engineering burden, the same rubric-and-question work a written or in-person exam requires. In our deployments, authoring a new exam and rubric took roughly four to eight hours, and we would not expect more than a day or two even for a wholly new discipline. Much of that authoring is itself AI-assistable, shared across a course's other assessments, and amortized over every exam that reuses it (Appendix~\ref{app:cost}).

\subsection{Grading Reliability}
\label{sec:reliability}

Initial independent grading in Fall 2025 revealed an inconsistency: Gemini averaged 16.3/20 while Claude averaged 13.0/20, a 3.3-point gap (GPT-5 fell between at 13.7/20, within one point of Claude on 64\% of students). Without deliberation, the council was useless. After deliberation, convergence was asymmetric: Gemini dropped its mean by 1.3 points after seeing stricter peer assessments, Claude rose by 0.9, GPT-5 by 0.3, all three converging toward the 14--15 range. That the largest shift came from the most lenient model suggests substantive revision rather than score-splitting.

\begin{figure}[tbp]
\centering
\includegraphics[width=0.7\columnwidth]{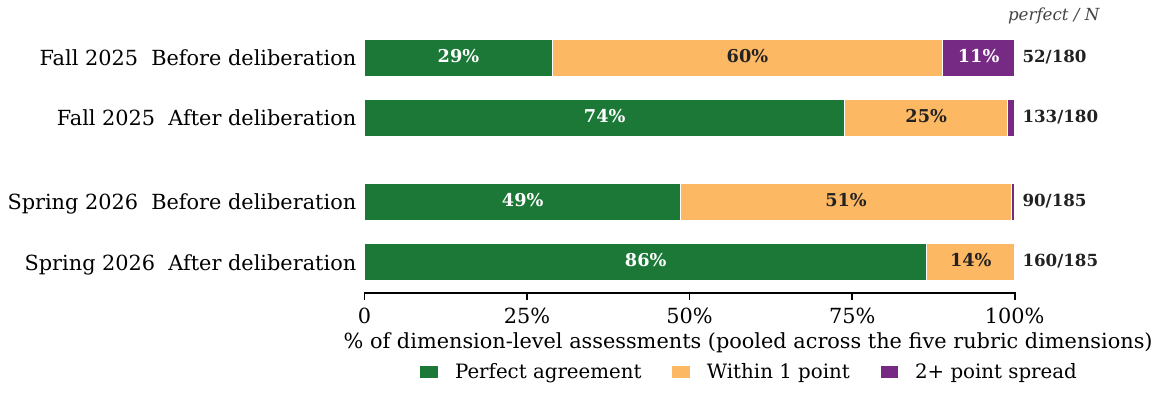}
\caption{Pooled grading agreement before and after deliberation. Each bar is the share of dimension-level assessments (pooled across the five rubric dimensions) where the three council models agreed perfectly (green), within one point (amber), or by 2+ points (purple). Fall 2025: $n=36$ (180 assessments); Spring 2026: $n=37$ (185). Per-dimension breakdown in Figure~\ref{fig:grading-agreement}.}
\Description{Four horizontal stacked bars showing pooled perfect/within-1/2+ agreement before and after deliberation for Fall 2025 and Spring 2026.}
\label{fig:grading-agreement-overall}
\end{figure}

\begin{figure}[tbp]
\centering
\includegraphics[width=0.82\columnwidth]{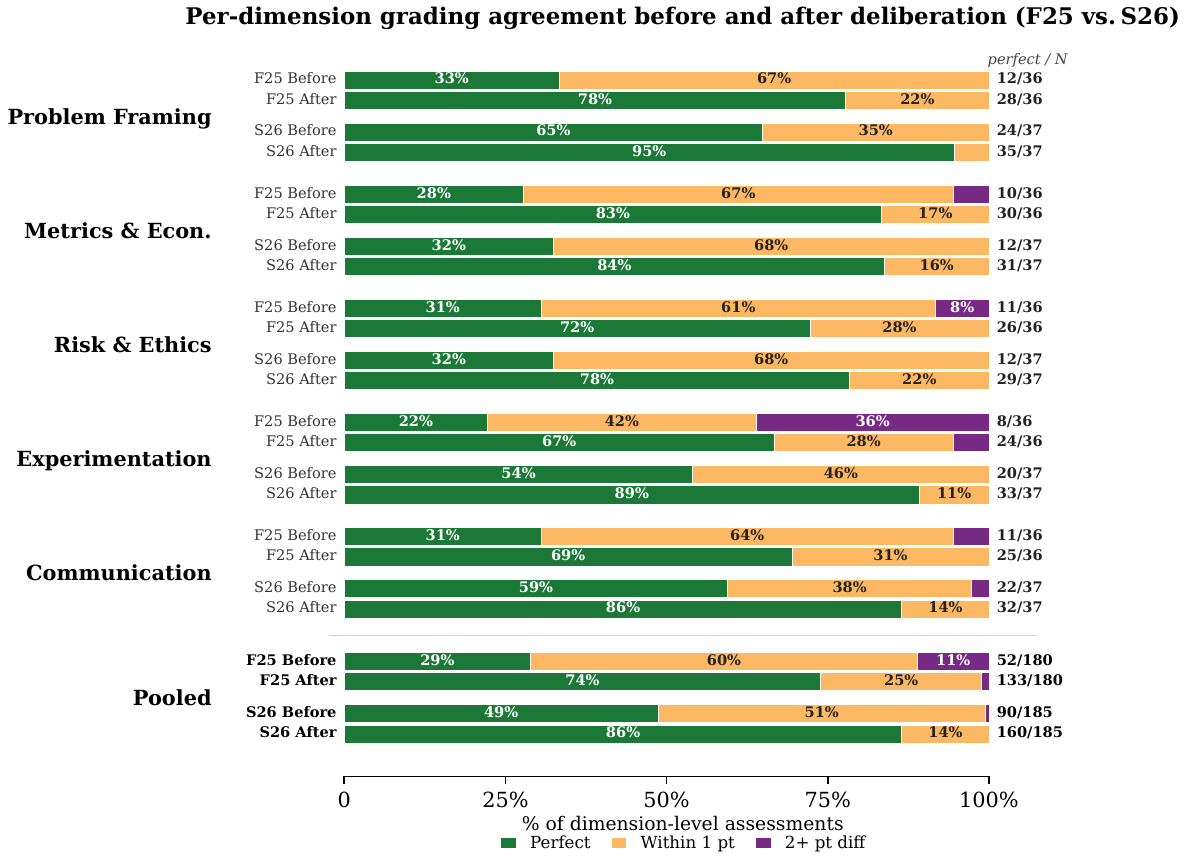}
\caption{Per-dimension grading agreement before and after deliberation, both cohorts (Fall 2025: $n=36$ students, 180 dimension-level assessments; Spring 2026: $n=37$ students, 185 assessments). Each row is the fraction of that cohort's assessments where the three council models agreed perfectly (dark green), landed within one point (amber), or diverged by 2+ points (dark purple). For each dimension, four bars are shown: F25 Before, F25 After, S26 Before, S26 After (the bottom block pools across all five dimensions). The right-hand column gives perfect-agreement count over total. The Before/After pair makes the deliberation gain visible directly: in F25, pooled perfect agreement rises 29\% $\to$ 74\% and 2+ point spreads collapse 11\% $\to$ 1\%; in S26 the same compression carries higher absolute agreement (49\% $\to$ 86\% perfect, with the single Round-1 2+ point spread also eliminated after deliberation). The dimension that ranks lowest on after-deliberation agreement shifts from Experimentation (F25, 67\% perfect) to Risk \& Ethics (S26, 78\%): noise patterns are not fixed properties of the rubric.}
\Description{Horizontal stacked-bar chart of per-dimension grading agreement for the two cohorts, showing both rounds of grading. Five dimensions on the y-axis, each with four bars (F25 Before, F25 After, S26 Before, S26 After); a pooled block at the bottom. Each bar split into perfect (green), within 1 (amber), 2+ (purple).}
\label{fig:grading-agreement}
\end{figure}

Deliberation improves agreement, with the largest relative gain in the noisier Fall 2025 cohort (Figure~\ref{fig:grading-agreement-overall}; per-dimension breakdown in Figure~\ref{fig:grading-agreement}). Perfect three-way agreement rose from 29\% to 74\% in Fall 2025 and from 49\% to 86\% in Spring 2026 (pooled across the five rubric dimensions); 2+ point spreads fell from 11\% to 1\% (Fall 2025) and from 1\% to 0\% (Spring 2026). We summarize agreement with Krippendorff's~$\alpha$, a chance-corrected agreement score on a $0$--$1$ scale: it discounts the agreement three raters would reach just by guessing, so it is a stricter bar than the raw percentages above. On that scale $\alpha \geq 0.80$ is the conventional threshold for good agreement~\cite{krippendorff2004content}; we compute $\alpha$ following~\cite{krippendorff2011}. Pooled the same way, $\alpha$ rose from $0.52 \to 0.86$ in Fall 2025 and $0.65 \to 0.90$ in Spring 2026. On the $0$--$20$ overall score it reached $0.83$ and $0.95$ after deliberation.

Pooled $\alpha$ above the 0.80 threshold is necessary but not sufficient: it does not say \emph{where} the residual disagreement lives. Decomposing the council shows that a single rater accounts for nearly all of it, and the bias is directional rather than random. In both cohorts Gemini grades systematically more lenient than the other two models on the 0--20 total (on average, Round~2 Gemini scored $+1.08$ points above Claude in Fall 2025 and $+0.38$ above in Spring 2026), while Claude and GPT-5 agree almost perfectly (pairwise $\alpha = 0.97$ in Fall 2025 and $0.99$ in Spring 2026; mean absolute total-score difference $\leq 0.4$ points in both). Dropping Gemini from the council would lift overall-score $\alpha$ to $0.97$ (Fall 2025) and $0.99$ (Spring 2026), not by reducing noise (the Claude--GPT-5 pair was already tight) but by removing the directional offset.

Deliberation compresses this offset but does not erase it. Round~1 Gemini entered the council roughly $+2$ to $+3.3$ points lenient on the 0--20 total, depending on cohort; in Round~2 it moved toward consensus without fully closing the gap. In Fall 2025 its lenient margin shrank from $+3.25$ to $+1.08$ points (${\sim}67\%$ of the gap absorbed), leaving roughly one third of it; in Spring 2026 the margin shrank from $+1.97$ to $+0.38$ (${\sim}81\%$ absorbed), leaving roughly one fifth. Per-model biases survive deliberation as stable offsets rather than average out.

We retain Gemini despite the apparent reliability gain from dropping it. A persistently lenient voice forces the stricter raters (Claude and GPT-5) to cite specific transcript evidence when defending deductions rather than reaffirm low scores by default. The practical lesson for LLM-as-judge ensembles: expect per-model biases to survive deliberation as stable offsets, and choose raters for argumentative asymmetry rather than interchangeability.

One objection to this argument is that the chair synthesizer is Claude, itself one of the two strict raters: if a strict chair can discard the lenient input, retaining Gemini may add less than the argumentative-asymmetry story implies. We do not have a controlled chair-swap experiment, but the chair output sits closer to the Round-2 mean than to either pole (chair-vs-Round-2 absolute difference $\leq 0.5$ points on the 0--20 total on average, with no systematic strict-side or lenient-side offset). That is consistent with the chair acting as a weighted synthesizer of all three Round-2 votes rather than a strict-side selector; a controlled alternative-chair comparison (Gemini or GPT-5 as chair, or a non-LLM aggregation rule) is left to future work.

The decomposition also yields a tighter audit trigger than raw three-way spread. Because residual disagreement concentrates in one rater with a stable directional offset, the most informative criterion is \emph{disagreement between the two tightly-coupled raters}. We flag an exam for human audit when Claude and GPT-5 themselves diverge by more than one point on the 0--20 scale, or when Gemini sits more than two points from the midpoint of Claude and GPT-5 (in either direction). The second clause is symmetric by design: although Gemini was consistently the lenient outlier in our two cohorts, an unexpected reversal (Gemini scoring well below the strict pair on some future cohort) would be at least as informative and should fire the same rule. Either condition signals disagreement that normal deliberation cannot absorb. In Spring 2026 this combined rule fires on roughly 3\% of exams (one student out of 37).

We treat human audit as a \emph{core} step of the grading protocol, not a post-hoc convenience: AI handles the volume, and the human handles the flagged minority and the cases the council itself surfaces as ambiguous. We did not measure audit-time-per-case; the property the rule guarantees is the flagged fraction (${\sim}3\%$ at the disagreement structure observed across two cohorts), not a per-case cost.

\subsection{Diagnostic Power for Teaching Quality}
\label{sec:validity}

Reliability (``do raters agree?'') is necessary but not sufficient; \emph{validity} (``do grades reflect actual student competence?'') requires expert human judgment. As a partial validity check, the instructor and teaching assistant independently graded all 36 Fall 2025 examinations and compared their scores to the council output. The instructor's holistic grades aligned with the most lenient model (Gemini), but after reviewing the stricter models' evidence (verbatim transcript excerpts supporting each deduction) the instructor conceded most of the stricter scores. Only two students requested regrading, and neither challenge succeeded; the transcript evidence backing each grade was too specific to dispute.

This comparison was informal, not a controlled validation study. The instructor's leniency was partly driven by awareness of teaching gaps, a confound we did not control for. Rigorous validation would require blinded expert re-grading of a random subset with pre-defined scoring criteria. Reliability gives us consistency among raters; the human auditor is the operational safeguard against validity failures, but formal validity (against blinded expert re-grading on pre-defined criteria) remains future work.

The same evidence trail benefits students directly: every exam yields a structured, evidence-linked feedback report as a byproduct of grading, something human graders rarely produce for every student (Section~\ref{app:discussion}).

\begin{figure}[tbp]
\centering
\includegraphics[width=0.8\columnwidth]{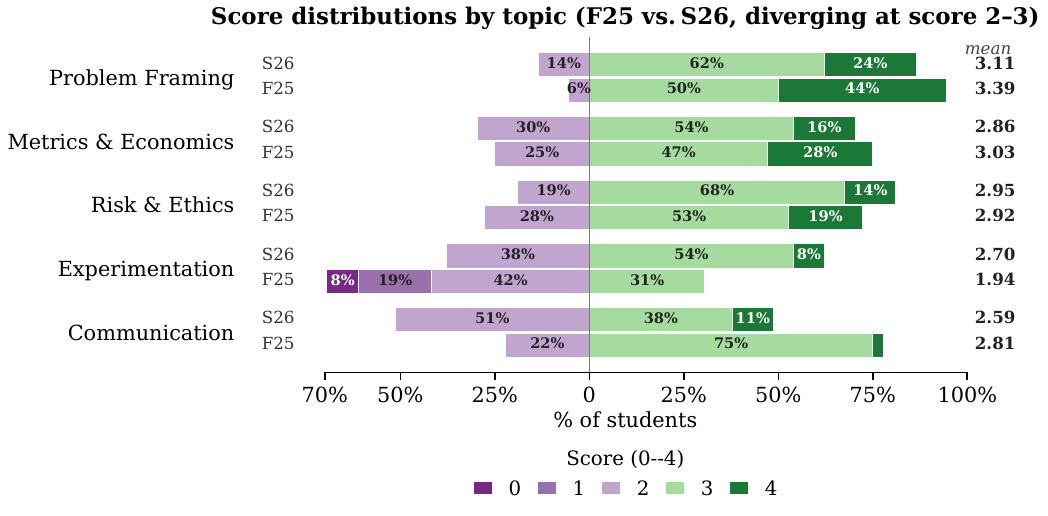}
\caption{Per-dimension score distributions across both cohorts (Fall 2025: $n=36$; Spring 2026: $n=37$). Scores 0--2 extend left of the centerline, 3--4 extend right; bold numbers are means. Fall 2025 Experimentation had a left tail (mean 1.94); Spring 2026 Experimentation shifted right (mean 2.70, no scores below 2) after we revised the syllabus to give A/B testing more class time.}
\Description{Horizontal diverging stacked bar chart with five rubric dimensions and two cohorts each.}
\label{fig:performance-by-topic}
\end{figure}

External grading has a second benefit over instructor self-grading: when you grade your own students, it is easy to give credit for what you \emph{know} they understood rather than what they \emph{demonstrated}, and the stricter council surfaces teaching deficiencies that would otherwise remain hidden. Scores by topic varied (Figure~\ref{fig:performance-by-topic}): Fall 2025 Experimentation (A/B testing) averaged 1.94/4 with a left tail (8\% scored 0, 19\% scored 1, zero demonstrated mastery), against Problem Framing at 3.39/4. This was a mirror held up to the instructors, not just the students: A/B testing had received insufficient class time, easy to overlook when instructors grade their own students leniently on topics they know were covered superficially. The gap closed in Spring 2026 after we rebalanced the syllabus: Experimentation rose to 2.70, with no student scoring below 2.

\begin{figure}[tbp]
\centering
\includegraphics[width=0.72\columnwidth]{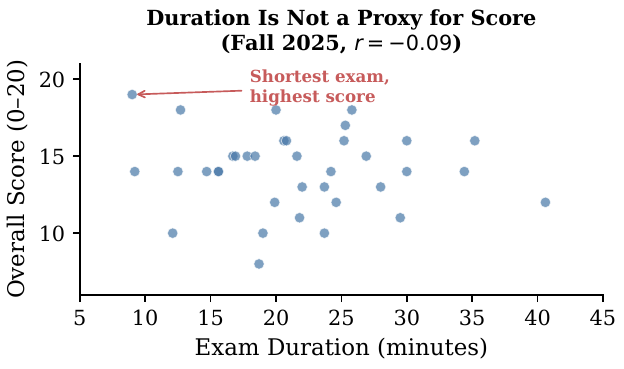}
\caption{Examination duration vs.\ overall score (Fall 2025, $n=36$). The shortest exam (9 minutes) received the highest score (19/20); the longest (41 minutes) received 12/20. Pearson $r = -0.09$ for Fall 2025 and $r = -0.30$ for Spring 2026 ($n=37$; not shown). Duration is not a reliable proxy for score in either cohort: short exams were not necessarily shallow, and long exams were not necessarily stronger.}
\Description{Scatter plot of exam duration on the x-axis (9--41 minutes) against overall score on the y-axis (8--19), showing little association.}
\label{fig:duration-vs-score}
\end{figure}

Duration was a poor proxy for score in both cohorts (Figure~\ref{fig:duration-vs-score}): the shortest Fall 2025 exam (9 minutes) earned the highest score (19/20), while the longest (41 minutes) earned one of the lowest (12/20). In Spring 2026 the trend was moderately negative.

\section{What Broke: What We Wish We Knew Before Deploying}
\label{sec:failure-modes}

Fall 2025 surfaced five failure modes: the agent stacked several questions into a single turn; it paraphrased questions when asked to repeat them; it treated a student's thinking pause as a dropped call; it selected case studies far from randomly; and, in a cloned professor's voice, it came across as harsh. Some we fixed in code for Spring 2026, some we could only mitigate at the prompt level, and one (silence as thinking) survived its fix. Each part below gives the problem, then what we changed, and for the surviving one, why it resisted.

\paragraph{Question Stacking}

The most damaging issue. The agent regularly fired off compound questions. From an actual transcript:
\emph{``Tighten it up for me: who is the user, and what decision do they make differently because of your product? And what is your North Star metric, plus one counter metric that might get worse if you over-optimize?''}
A Fall 2025 student summed it up: \emph{``It kept asking 3 to 4 questions at the same time, despite me asking it to slow down multiple times.''} Over half of the student improvement suggestions on the exit survey mentioned this problem. Prompt-level instructions helped but did not eliminate the behavior: for constraining LLM behavior in extended conversations, they are necessary but insufficient. The failure is not unique to our stack: Socratic Mind's authors observed the same multi-question drift and planned a prompt-level fix~\cite{hung2024socratic}. We also built an ``interference protocol'' into the grading rubric, instructing graders to assess only the answered parts when the agent stacked questions and the student answered only some.

The lesson: \emph{enforce behavioral constraints through system architecture, not just prompting.} Spring 2026 added a programmatic turn validator that detects a multi-question turn and asks the model to re-emit a single-question version. A stacked turn packs two or more separate questions into a single agent turn; their rate fell from 31\% in Fall 2025 to 11\% in Spring 2026.\footnote{Precisely, 31.2\% of agent question-turns (191 of 612) fell to 10.9\% (122 of 1{,}115). The drop also holds exam by exam (the most-stacked exam fell from 85\% to 39\%).} The turns that remained also got simpler: three-or-more-question turns nearly vanished, falling from 2.9\% to 0.3\%; questions spanning two rubric dimensions roughly halved, from 10.6\% to 4.6\%; and the median stacked turn shrank from 52 to 38 words. A few three- and four-question turns still surface, so this is a large reduction, not elimination.

\paragraph{Paraphrasing During Clarification}

When students asked the agent to repeat a question, it would paraphrase rather than repeat verbatim, inadvertently changing the question. The LLM was doing what it was trained to do (being helpful by rephrasing) but exactly the wrong reflex in an examination. Explicit prompt instructions (``repeat the exact previous question'') reduced but did not eliminate the behavior; this remains a prompt-level mitigation, not a code-enforced one.

\paragraph{Silence Handling}

By default the platform treated a silence as a sign that something had gone wrong (a dropped call) rather than as a student thinking. Fall 2025 ran on the 5-second default, so the agent would cut in with ``Are you there?'' exactly when a student was formulating an answer. Spring 2026 raised the threshold to 15 seconds, but the interruptions persisted, and students reported that the check-ins spiked their stress precisely when they were trying to think. The lesson: silence is thinking time. Our default for future semesters is 30 seconds.

\paragraph{Non-Random Case Selection}

\begin{figure}[tbp]
\centering
\includegraphics[width=0.8\columnwidth]{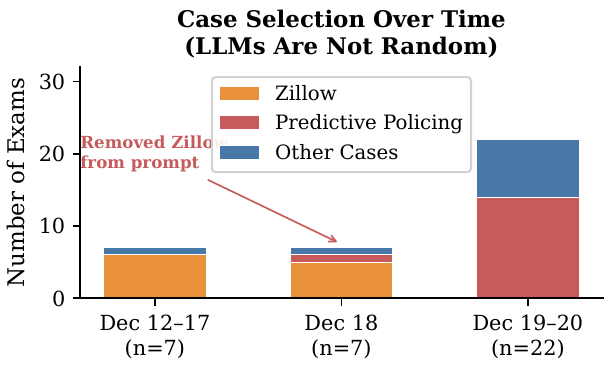}
\caption{Fall 2025 case selection over time. Despite instructions to ``randomly select'' a case from a library of eight, the agent chose Zillow in 6 of 7 examinations during December 12--17 (86\%). After Zillow was removed from the prompt at the end of December 18, the agent fixated on Predictive Policing in 14 of 22 examinations on December 19--20 (64\%). Spring 2026 (not shown): a deterministic seed-to-case mapping in code produced a 22\% maximum share with all eight cases used.}
\Description{Stacked bar chart of three time periods showing case selection skew: Zillow dominates Dec 12--17, Predictive Policing dominates Dec 19--20.}
\label{fig:case-selection}
\end{figure}

This was the most instructive failure (Figure~\ref{fig:case-selection}). The examination prompt listed eight case studies and instructed the agent to ``randomly select ONE case.'' From December 12--17, when the Zillow case study was on the list, the agent selected it in 6 of 7 examinations (86\%). Zillow's iBuyer home-buying program was the case used to discuss concept drift and adverse selection. At the end of December 18, we removed this case study from the prompt. The agent immediately shifted to the Predictive Policing case study, selecting it for 14 of 22 examinations on December 19--20 (64\%), better than the level of fixation on the Zillow case study but still far from the ${\sim}14\%$ expected under uniform selection over the remaining seven cases. A predictable case is also a leakable one: a student who can anticipate which case they will receive can prepare that scenario in advance or tip off the next student, undercutting the leak-resistance that the dynamically generated questions otherwise give the format (Section~\ref{sec:cannot-do}).

The fundamental issue: LLMs should not be trusted to implement uniform randomization by instruction. The same deterministic bias appears when models are asked simply to generate random numbers~\cite{coronado2025}. They carry ordering biases that survive every prompt-level mitigation we tried, and the bias is invisible until you examine the output distribution. Kelly~\cite{kelly2026randomization} makes the broader case that randomization is routinely botched even when practitioners are explicitly trying to implement it; asking an LLM to randomize by instruction is a textbook instance. The fix is to draw a random number in code and map it to a case deterministically, which is what Spring 2026 shipped. Across the 37 Spring 2026 examinations, all eight cases were used and the maximum single-case share was 22\% (8/37), in contrast to the Fall 2025 single-week 86\% peak.\footnote{With $n = 37$ across 8 cases, expected cell counts under uniformity are 4.6, below the conventional ${\geq}5$ rule for the $\chi^2$ approximation, so we report the observed distribution descriptively rather than as a goodness-of-fit test against uniformity.} The principle generalizes: any behavior requiring uniformity or randomness belongs in code, not in the LLM.

\paragraph{Voice Characteristics}

We cloned the voice of a course professor from a conference keynote, reasoning that a familiar voice would feel more natural than the defaults. The opposite happened: students described the agent as ``aggressive,'' ``condescending,'' ``shouting,'' a ``sports coach'' style ``absolutely not appropriate for classroom use.'' One free-response answer reduced it to four words: \emph{``Make it less mean.''} Keynote delivery (fast-paced, projected, designed to talk \emph{at} a large audience) is the wrong register for a one-on-one exam; without the softening cues of face-to-face interaction, a high-fidelity clone of an authority figure read as intimidating rather than familiar. Spring 2026 switched to ``Alice,'' an ElevenLabs preset with a calm, education-oriented register; the change removed the complaints. Voice characteristics warrant the same attention as question design.

Table~\ref{tab:design-patterns} collects the five patterns, pairing each with the failure it addresses and the implementation that enforced the fix; paraphrasing remains the one failure without an architectural pattern, mitigated only at the prompt level.

\begin{table}[t]
\caption{Design Patterns for Voice AI Assessment Systems.}
\label{tab:design-patterns}
\footnotesize
\centering
\begin{tabular}{>{\raggedright\arraybackslash}p{0.21\columnwidth}>{\raggedright\arraybackslash}p{0.30\columnwidth}>{\raggedright\arraybackslash}p{0.40\columnwidth}}
\toprule
\textbf{Pattern} & \textbf{Failure Addressed} & \textbf{Implementation} \\
\midrule
1. Decompose into single-purpose modules &
A monolithic do-everything agent; conversational drift across exam phases; cascading prompt failures &
Separate agents per phase (authentication, project, case); independent grading and chair-synthesis steps; phase transitions enforced by orchestrator, not LLM discretion \\
\midrule
2. Constrain in code or configuration, not prompts &
Question stacking (code); silence handling (platform configuration) &
Programmatic turn validator rejects multi-question responses; voice-activity-detection timeout raised from 5\,s to 15\,s (itself insufficient; 30\,s is the future default) \\
\midrule
3. Never delegate randomization to the LLM &
Non-random case selection (ordering bias survives every prompt) &
Deterministic case mapping from a per-session random seed passed as a dynamic variable \\
\midrule
4. Multi-model council with deliberation &
Single-model grading bias; silent miscalibration &
Independent scoring (Round~1), cross-review with cited evidence (Round~2), chair synthesis; flag high-disagreement cases for human audit \\
\midrule
5. Voice selection &
Perception of aggressiveness &
Selected a calm, education-oriented voice \\
\bottomrule
\end{tabular}
\end{table}

\section{Student Experience}
\label{sec:student-experience}

The central tension held in both cohorts: what the exam measured well versus how it felt to take. Fall 2025 students rated ``tested my actual understanding'' highest (70\% agreement), but only 33\% found questions clear and 83\% found the exam more stressful than written exams. Spring 2026 replicated the student-perceived-understanding finding (66\%) while clarity rose to 78\% and ``more stressful'' fell to 63\%; the timing coincides with the between-cohort engineering fixes, though with several changes at once we cannot say which fix moved which response. ``Tested my actual understanding'' is a student-perceived measure, not a psychometric validity claim; the reliability-vs-validity distinction is in Section~\ref{sec:validity}.

\begin{table}[htbp]
\caption{AI Examiner Experience, Fall 2025 (5-point agree--disagree scale, $n=30$)}
\label{tab:survey-likert}
\small
\begin{tabular}{lrrrrrc}
\toprule
\textbf{Item} & \textbf{SD} & \textbf{D} & \textbf{N} & \textbf{A} & \textbf{SA} & \textbf{\% Agree} \\
\midrule
Questions clear and understandable & 4 & 13 & 3 & 10 & 0 & 33\% \\
Enough time to think & 7 & 8 & 3 & 9 & 3 & 40\% \\
Felt like a fair evaluator & 3 & 9 & 8 & 8 & 2 & 33\% \\
Comfortable speaking to AI & 5 & 11 & 7 & 5 & 2 & 23\% \\
Conversation flowed naturally & 4 & 11 & 4 & 10 & 1 & 37\% \\
Tested actual understanding & 2 & 6 & 1 & 17 & 4 & 70\% \\
Voice interaction worked well & 0 & 2 & 11 & 17 & 0 & 57\% \\
\bottomrule
\end{tabular}
\par\smallskip
\footnotesize Note: SD=Strongly Disagree, D=Disagree, N=Neutral, A=Agree, SA=Strongly Agree
\end{table}

\begin{table}[htbp]
\caption{AI Examiner Experience, Spring 2026 (5-point agree--disagree scale, $n=32$)}
\label{tab:survey-likert-s26}
\small
\begin{tabular}{lrrrrrc}
\toprule
\textbf{Item} & \textbf{SD} & \textbf{D} & \textbf{N} & \textbf{A} & \textbf{SA} & \textbf{\% Agree} \\
\midrule
Questions clear and understandable & 1 & 2 & 4 & 19 & 6 & 78\% \\
Enough time to think & 3 & 5 & 6 & 12 & 6 & 56\% \\
Felt like a fair evaluator & 0 & 4 & 10 & 15 & 3 & 56\% \\
Comfortable speaking to AI & 2 & 9 & 7 & 10 & 4 & 44\% \\
Conversation flowed naturally & 0 & 9 & 6 & 13 & 4 & 53\% \\
Tested actual understanding & 1 & 4 & 6 & 17 & 4 & 66\% \\
Voice interaction worked well & 0 & 3 & 7 & 22 & 0 & 69\% \\
\bottomrule
\end{tabular}
\par\smallskip
\footnotesize Note: SD=Strongly Disagree, D=Disagree, N=Neutral, A=Agree, SA=Strongly Agree
\end{table}

These rates come from voluntary surveys run after the exam but before grade release (30 of 36 responding in Fall 2025, 32 of 37 in Spring 2026; the instrument and full response distributions are in Appendices~\ref{app:survey-instrument} and~\ref{app:survey-results}). Such surveys can over-represent polarized reactions, so we checked the one thing we observe for non-respondents, their final grades: in both cohorts they were statistically indistinguishable from respondents' (Appendix~\ref{app:nonresponse}).

\paragraph{Prior oral-exam experience does not appear to explain the stress response.}
A natural reading of the headline rate (83\% Fall 2025 / 63\% Spring 2026 finding the AI exam more stressful than written exams) is a familiarity effect: 83\% (Fall 2025) and 88\% (Spring 2026) of respondents had never taken an oral exam in any course. Table~\ref{tab:crosstab-q2-q10} does not support this reading. Every respondent with prior oral-exam experience (all 5 in Fall 2025 and all 4 in Spring 2026) reported the AI exam more stressful, whereas among students without prior experience the ``more stressful'' rate was 80\% (Fall 2025) and only 57\% (Spring 2026). If novelty were driving the response we would expect the two rates to track; instead, the share of students who had never taken an oral exam held steady across cohorts (83\%\,$\to$\,88\%) while the stress rate moved (83\%\,$\to$\,63\%) with the engineering changes (Section~\ref{sec:failure-modes}). The prior-experience cells are small ($n = 5$ and $n = 4$) and the comparison is observational rather than experimental, so this is a claim about what does \emph{not} explain the observed pattern rather than a positive identification of the causal driver; a firmer test will require more cohorts.

\paragraph{Format preference is more consistent with perceived stress than with prior oral-exam experience.}
Format preferences also shifted between cohorts. Fall 2025: 57\% preferred traditional written exams, 27\% preferred a human-administered oral exam, 13\% preferred an AI-administered oral exam. Spring 2026: 53\% preferred written (stable), 19\% preferred human-oral (down from 27\%), and 25\% preferred AI-oral (up from 13\%). The apparent Fall 2025 gap (27\% preferring a human oral exam despite only 17\% having ever taken one) did not replicate as an experience-novelty effect: prior oral-exam experience did not consistently predict human-oral preference (3 of 8 in Fall 2025, 0 of 6 in Spring 2026; Table~\ref{tab:crosstab-q12}). What did replicate was a stress gradient across preferences. Students preferring written exams almost unanimously reported the AI exam as more stressful (17/17 in Fall 2025, 16/17 in Spring 2026); students preferring an AI oral exam did so less often than written-preferrers and less often in Spring 2026 (2/4 in Fall 2025, 1/8 in Spring 2026). The shift from ``prefer human oral'' to ``prefer AI oral'' is therefore more consistent with the drop in perceived stress than with prior oral-exam experience.

Two caveats temper that reading. Comfort and fairness ratings also rose between cohorts (Tables~\ref{tab:survey-likert}, \ref{tab:survey-likert-s26}), so a simultaneous shift in trust in AI cannot be ruled out as a contributor; the cross-tab pattern is informative about stress but does not isolate it from concurrent trust-in-AI effects. The shift coincides with the failure-mode fixes (Section~\ref{sec:failure-modes}). That 44\% of Spring 2026 respondents were open to an oral format (whether AI- or human-administered), up from 40\% in Fall 2025, suggests the remaining failure modes (not a categorical objection to oral assessment) are what gates further acceptance.

\begin{table}[htbp]
\caption{Prior oral-exam experience (Q2) $\times$ stress vs.\ written (Q10). Row percentages.}
\label{tab:crosstab-q2-q10}
\small
\begin{tabular}{llrrrr}
\toprule
\textbf{Cohort} & \textbf{Prior oral?} & \textbf{More} & \textbf{Same} & \textbf{Less} & \textbf{N} \\
\midrule
Fall 2025   & Never taken & 20 (80\%)  & 1 (4\%)  & 4 (16\%) & 25 \\
            & Had taken   & 5 (100\%)  & 0        & 0        & 5  \\
\midrule
Spring 2026 & Never taken & 16 (57\%)  & 5 (18\%) & 7 (25\%) & 28 \\
            & Had taken   & 4 (100\%)  & 0        & 0        & 4  \\
\bottomrule
\end{tabular}
\par\smallskip
\footnotesize ``More''/``Less'' pool ``much'' + ``a bit'' responses.
\end{table}

\begin{table}[htbp]
\caption{Format preference (Q12) $\times$ prior oral experience (Q2) and stress (Q10). Cells are counts; percentages are row shares of $N$.}
\label{tab:crosstab-q12}
\small
\begin{tabular}{llrrrr}
\toprule
\textbf{Cohort} & \textbf{Preference} & \textbf{Prior} & \textbf{More stressful} & \textbf{N} \\
\midrule
Fall 2025   & Written exam                  & 2  (12\%) & 17 (100\%) & 17 \\
            & Human oral exam               & 3  (38\%) & 6  (75\%)  & 8  \\
            & AI oral exam                  & 0          & 2  (50\%)  & 4  \\
            & No preference                 & 0          & 0          & 1  \\
\midrule
Spring 2026 & Written exam                  & 3  (18\%) & 16 (94\%)  & 17 \\
            & Human oral exam               & 0          & 2  (33\%)  & 6  \\
            & AI oral exam                  & 1  (13\%) & 1  (12\%)  & 8  \\
            & No preference                 & 0          & 1  (100\%) & 1  \\
\bottomrule
\end{tabular}
\par\smallskip
\footnotesize ``Prior'' = Q2 == True (had taken an oral exam in some course before); ``More stressful'' pools ``much'' + ``a bit'' more-stressful responses on Q10.
\end{table}

The most praised feature in both cohorts was scheduling flexibility. The most common Fall 2025 complaint was question stacking, matching the 31\% stacked-turn rate (Section~\ref{sec:failure-modes}); the complaint receded in Spring 2026 free responses as the rate dropped to roughly 11\%.

Accessibility cut both ways. Some Fall 2025 students with anxiety-related accommodations volunteered that they found the AI examiner less stressful than typical exams, suggesting the format could serve as a test-anxiety accommodation in some populations. For international students, the format cut the other way: ``At times, the difficulty wasn't in knowing the material but in articulating my thoughts under pressure in English.'' We added a live transcript display for Spring 2026 in response to Fall 2025 requests; it lets students catch mistranscriptions in real time, but cognitive load under pressure remains (caveats and a fuller accessibility discussion in Section~\ref{app:discussion}).

One complaint increased in Spring 2026: the adaptive nature of questioning was perceived as ``unfair''; students want exam difficulty to be the same for everyone. Asked to agree or disagree that the examiner ``felt like a fair evaluator,'' 33\% agreed in Fall 2025 and 56\% in Spring 2026. Yet after grades were released, only two regrading requests came in, both Fall 2025, and no points were conceded in either case. The complaint was about process, not outcome.

\section{Discussion}
\label{sec:discussion}
\label{sec:cannot-do}

We now weigh what the system means for anyone considering adoption: what AI administration actually buys, what the grades can and cannot claim, the academic-integrity and accessibility trade-offs, and how student data is handled.

\subsection{What AI Administration Buys}
\label{app:ai-buys}

Human audit is a core step of the protocol rather than an optional overlay, as Section~\ref{sec:reliability} concedes. A fair question then follows: \emph{if scoring still requires instructor validation, what is gained?} Three things, each scaling differently with cohort size.

\paragraph{Automation of the time-dominant interview phase.}
At our per-exam durations (22-minute mean in Fall 2025, nominal 30 minutes in Spring 2026) and cohort sizes, the single-examiner interview time alone would consume roughly 13 hours in Fall 2025 and 19 in Spring 2026 ($36 \times 22 / 60$ and $37 \times 30 / 60$); doubling for a two-grader configuration at the same actual durations gives about 26 and 37 hours. Under the nominal 25-minute, two-grader comparator of Appendix~\ref{app:cost}, the same cohorts would land at roughly 30 and 31 hours. Either framing excludes scheduling, room booking, and between-exam overhead, and would serialize along the examiners' calendars. The AI absorbs those hours concurrently. The interview phase is the time-dominant cost of traditional orals, and it is the phase the AI fully replaces.

\paragraph{Audit load scales with the flagged fraction rather than requiring universal human review.}
The decomposition-informed audit rule fired on roughly 3\% of Spring 2026 exams: one student out of 37, and on the order of 30 flagged exams at 1{,}000 students if the rate holds. The absolute count still grows with the cohort; the operational saving is that the human reviews a small fraction rather than every exam.

\paragraph{Scale.}
At 200, 500, or 1{,}000 students a faculty-administered oral is infeasible for most courses without a dedicated corps of trained graders. So the realistic counterfactual (what would otherwise happen) at those sizes is not AI-vs-human but AI-vs-written-only, with the integrity weaknesses that motivated this paper. The point of comparison shifts from ``can the AI match a faculty oral'' to ``does an AI oral provide assessment properties that the only feasible alternative (a written exam) cannot.''

\subsection{What the Grades Can and Cannot Claim}

The caveats start with what the grades mean. The grading council's high agreement establishes reliability (consistency among AI raters), not validity (correctness against expert judgment); the human auditor on roughly 3\% of Spring 2026 exams is a guardrail, not a full fix. What we can claim is operational feasibility and promising reliability. What we cannot claim is that these grades carry the validity of a well-run faculty oral, or that the format is demonstrably fair to every student: the first would take blinded expert re-grading, the second a subgroup fairness analysis (checking whether scores hold up evenly across student groups), and we have performed neither.

\subsection{Academic Dishonesty Vectors}
\label{app:cheating}

The format inverts the traditional cheating incentive. The LLM generates questions dynamically against published guidelines rather than from a fixed question bank, so there is no exam to leak. Practicing against the published structure is itself the learning behavior the format rewards.

Three forms of academic dishonesty deserve explicit treatment given the unsupervised home environment in which the exams ran.

\paragraph{Pre-rehearsed answers.}
Dynamic follow-up questioning exposes the limits of rehearsal: questions are calibrated to what the student just said, so a scripted answer that does not reflect actual understanding breaks down within one or two follow-ups. The adversarial edge case collapses into the best-case scenario. A student who reverse-engineers the examiner prompt and rehearses exhaustively has, for all practical purposes, practiced explaining the material out loud until they can answer arbitrary follow-up questions: they have learned it.

\paragraph{Live AI coaching.}
A second AI system feeding answers in real time via headset is not fully preventable: a fluent student repeating coached answers naturally would not be detectable from the transcript or video alone. This is a broader societal problem, not specific to this system: the same concern applies to human job interviews and is already surfacing with audio earpieces and camera-equipped glasses. No mitigation we are aware of is absolute.

\paragraph{Confederate in the room.}
A second person whispering or typing answers off-camera is deterred by the self-recorded video requirement, which instructors can review. Detection is not guaranteed (camera placement, off-screen displays, and lip-read coaching all remain possible), but the marginal effort and risk of detection are higher than in the unproctored written setting.

\paragraph{Net.}
None of these mitigations are absolute. The format is more integrity-preserving than unproctored written work and does not overclaim: we do not claim to prevent all cheating, only to raise the effort required above the level where AI assistance provides a free ride.

\subsection{Feedback, What Is Lost, and Accessibility}
\label{app:discussion}

\paragraph{Feedback Quality.}
The grading council produced feedback exceeding what human graders typically deliver in both structure and specificity. Each evaluation included strengths, weaknesses, and action items with verbatim evidence from the transcript. For the highest-scoring student: ``Your understanding of metric trade-offs and Goodhart's Law risks was exceptional; the hot tub example perfectly illustrated how optimizing for one metric can corrupt another.'' For a lower-performing student: ``Practice articulating complete A/B testing designs: state a hypothesis, define randomization unit, specify guardrail metrics, and establish decision criteria for shipping or rolling back.'' Human graders rarely generate such detailed, evidence-linked feedback for every student. The automated system produces it as a byproduct.

\paragraph{What Is Lost.}
Automating assessment is not a free lunch. A human examiner develops real-time intuition about student understanding, noticing by the third or fourth exam which topics the class systematically struggles with. Our Experimentation gap (Section~\ref{sec:validity}) was only visible \emph{after} analyzing aggregate scores from the AI system; a human examiner might have noticed it sooner. But this concern can be mitigated by other feedback channels. In our course, students complete a brief exit survey after every session (what they learned, already knew, and found unclear), and an LLM analyzes responses to generate pedagogical recommendations within minutes. We apply the same approach to assignments before grading, identifying class-wide patterns early enough to address in subsequent sessions. These mechanisms are orthogonal to the exam format and provide the continuous pedagogical feedback that automated assessment alone lacks. There is also value in the human relationship between examiner and student. A professor can read body language, offer encouragement, and exercise contextual judgment that no rubric fully captures. We do not claim AI-administered assessment is superior to well-resourced human assessment. The claim is narrower: it is superior to \emph{no} oral assessment, the realistic alternative for most courses.

\paragraph{Accessibility.}
Voice-based assessment introduces accessibility challenges that institutions must address before deployment, not after. Students with speech-related disabilities or conditions affecting verbal fluency face barriers that written formats avoid. For non-native speakers, the picture is more complicated than it might appear: because speech-to-text output feeds an LLM rather than a keyword matcher, the system can in principle apply domain-aware contextual repair that standalone automatic speech recognition (ASR) systems cannot (we did not benchmark this against a word error rate (WER) baseline, so we report it as a mechanism that may partly offset accented-speech errors, not as a measured reduction). Pre-exam, several Fall 2025 students worried the system would not understand their accent; post-exam, the barrier proved cognitive rather than acoustic (the international student quoted in Section~\ref{sec:student-experience}).

This points to a distinct fairness risk: a live oral rewards fast verbal production, so a student who understands the material but formulates answers more slowly (because of a processing-speed difference, a stutter, or simply thinking in a second language) can be underserved by a format that a written exam would not penalize. The silence-timeout fix (Section~\ref{sec:failure-modes}) partly addresses this by not punishing thinking pauses, but pacing pressure remains. The accommodations we consider necessary before wider deployment therefore include configurable (not merely extended) time limits and pacing, a choice of agent voice and speaking rate, an ungraded practice run so the interface is familiar before stakes are real, and a text-based alternative for students who cannot use the voice format; accented speech in particular should be spot-checked against gold transcripts rather than assumed handled. Spring 2026 added a live transcript display, a direct response to Fall 2025 exit-survey requests, giving students real-time confirmation of what the system understood and allowing immediate rephrasing. Students should be informed in advance about how the system works, what is recorded, and who accesses their transcripts. The format is not uniformly exclusionary: students with anxiety-related accommodations found the AI examiner \emph{less} stressful than facing a senior professor, and the live transcript particularly benefits non-native speakers and those using technical vocabulary, suggesting voice AI may simultaneously create and remove accessibility barriers depending on the population.

\subsection{Data Governance}

The system transmits voice audio and transcripts to ElevenLabs and graded-transcript text to Anthropic, Google, and OpenAI for council grading, governed by each provider's API privacy policy and (where available) by enterprise/zero-retention agreements; we did not run an independent audit. Locally recorded webcam and audio (a course-integrity requirement, Section~\ref{sec:results}) are not transmitted to any vendor.

\section{What Transfers, and What Doesn't}
\label{app:replicated}

The Scope-and-contribution paragraph at the end of Section~\ref{sec:problem} distinguishes patterns that replicated across the two cohorts from numbers that are cohort-specific instances. The full split is given in Table~\ref{tab:replicated} and discussed below.

\begin{table}[h]
\footnotesize
\caption{Replicated structure vs.\ cohort-specific magnitude across Fall 2025 and Spring 2026. Replicated rows are what we offer as the transferable Pattern~4 result; cohort-specific rows are instances rather than estimates of a population.}
\label{tab:replicated}
\centering
\begin{tabular}{>{\raggedright\arraybackslash}p{0.32\columnwidth}>{\raggedright\arraybackslash}p{0.28\columnwidth}>{\raggedright\arraybackslash}p{0.30\columnwidth}}
\toprule
\textbf{Property} & \textbf{Fall 2025} & \textbf{Spring 2026} \\
\midrule
\multicolumn{3}{l}{\emph{Replicated (direction and shape)}} \\
\midrule
Gemini enters lenient (Round 1 Gemini $-$ Claude on 0--20) & $+3.25$ pts & $+1.97$ pts \\
Gemini compresses after deliberation & $+3.25 \to +1.08$ pts ($\sim$$\frac{2}{3}$ of gap absorbed) & $+1.97 \to +0.38$ pts ($\sim$$\frac{4}{5}$ of gap absorbed) \\
Claude--GPT-5 tight pair (Round 2 pairwise $\alpha$, overall) & $0.97$ & $0.99$ \\
Residual disagreement concentrates in one rater (drop-Gemini leave-one-out $\alpha$) & $0.97$ & $0.99$ \\
\midrule
\multicolumn{3}{l}{\emph{Cohort-specific (magnitude)}} \\
\midrule
Round 2 Krippendorff's $\alpha$ (dim-pooled, ordinal) & $0.86$ & $0.90$ \\
Round 2 Krippendorff's $\alpha$ (overall, ordinal) & $0.83$ & $0.95$ \\
Lowest-agreement rubric dimension & Experimentation (67\% perfect) & Risk \& Ethics (78\% perfect) \\
Magnitude of residual Gemini offset (Round 2, signed mean on 0--20) & $+1.08$ pts & $+0.38$ pts \\
Marginal grading-LLM cost per exam & ${\sim}\$0.29$ & ${\sim}\$0.96$ \\
Marginal cost per student-minute (within subscription) & ${\sim}\$0.01$ & ${\sim}\$0.03$ \\
\bottomrule
\end{tabular}
\end{table}

\paragraph{What we offer as transferable.}
Three regularities replicated in direction across the two cohorts: (1) Gemini consistently entered Round~1 lenient on the 0--20 total; (2) Claude and GPT-5 behaved as an interchangeable tight pair at pairwise $\alpha > 0.96$; (3) residual disagreement concentrated in a single rater with a directional offset rather than distributing as random noise. These are the empirical basis for Pattern~4 (use a multi-model council with raters chosen for argumentative asymmetry).

\paragraph{What we do not claim transfers.}
The absolute $\alpha$ values, the identity of the lowest-agreement rubric dimension, the magnitude of the residual Gemini offset, and the per-student dollar figures are instances from two cohorts of one course at one institution. Two iterations cannot distinguish a two-cohort coincidence from a general property; we report the per-cohort numbers as instances rather than estimates of a population. The patterns themselves are stated in terms of LLM properties rather than classroom properties, which is why we expect them to generalize, but that expectation is a conjecture rather than a demonstrated result.

\paragraph{Adaptation cost as a scope boundary inside higher education.}
Even within higher education our deployment is a single course format (a capstone project plus case discussion) in a single discipline (AI/ML product management). The adaptation cost flagged in Appendix~\ref{app:cost} (rubric, case library, phase-specific agent prompts, pilot runs) would be paid again before any of the numbers above could be re-measured on a new population. Nothing in our experience predicts how that adaptation cost scales with disciplinary distance from the original deployment.

\paragraph{Plausible next test sites.}
Neighboring domains where analogous structured spoken evaluation exists are the most plausible next test sites for the patterns: technical hiring phone screens, where Jabarian and Henkel~\cite{jabarian2025} report AI interviews increasing job offers by 12\% and retention by 16--18\% across 70{,}000 applicants; and professional certification formats (medical boards, legal competency, regulatory compliance). We have not deployed in any of them, and we do not claim the specific numbers would survive the move.

\section{What's Next}
\label{app:whats-next}

Four extensions seem most immediately valuable.

\paragraph{Retrieval-augmented examination.}
In both Fall 2025 and Spring 2026 the agent received project summaries as parameters but could not access student-submitted artifacts. If it could reference specific slides or code, it could ask ``On slide 7, you claim 94\% accuracy. Walk me through how you validated that number.'' This would further close the gap between artifact quality and demonstrated understanding.

\paragraph{Formative deployment.}
The same infrastructure could support low-stakes practice examinations throughout the semester, letting students test understanding before high-stakes assessment. At a marginal cost on the order of \$1 per 10 minutes of graded exam time in the above-subscription regime (Section~\ref{sec:cost}), a more immediate application is attaching a short oral comprehension check to every written assignment, a pass/fail conversation verifying that the student can explain what they submitted. It converts the system from a semester-end assessment tool into a standing check attached to every assignment.

\paragraph{Validated expert re-grading of the flagged subset.}
The decomposition-informed audit rule already in place fires on roughly 3\% of Spring 2026 exams (Section~\ref{sec:reliability}), and instructor + TA review of those flagged cases is now treated as a core step of the protocol. The next step is a controlled validation study (blinded expert re-grading of a random subset with pre-defined scoring criteria), so the gap from inter-rater reliability to validity against expert judgment can be measured rather than asserted.

\paragraph{Accessibility by design.}
Spring 2026 shipped the live transcript display. The accommodations enumerated in Section~\ref{app:discussion} (configurable time limits and pacing, a choice of agent voice and speaking rate, an ungraded practice run, and a text-based alternative for students who cannot use voice) remain to be built, and moving the suggestive accessibility observations there from preliminary to validated would need a systematic study of students with disabilities, which we have not conducted.

\section{Conclusion}
\label{sec:conclusion}

The spread of LLMs has disrupted traditional assessment. Rather than pursuing a detection arms race, we can adopt formats that inherently reward actual understanding. Oral examinations have always had this property but never scaled; voice AI restores scalability. Our deployment demonstrates operational feasibility across two cohorts (73 students). Agreement within the grading council is high. That is encouraging, but it does \emph{not} by itself settle validity against double-blind expert human graders. Student experience was mixed. Students felt the exams tested their understanding but found them more stressful than written exams; the ``more stressful'' rate fell from the first cohort to the second as engineering fixes shipped, and question stacking, the most-cited culprit, fell with it. They dislike the adaptive nature of questioning as procedurally unfair; yet once grades were released, almost no one disputed them (Section~\ref{sec:student-experience}).

The practical contribution is the five patterns collected in Table~\ref{tab:design-patterns}. Two iterations were enough to surface them; how far they carry is what we hope future deployments will show. The same infrastructure has an application beyond the final exam: attaching a short oral comprehension check to every written assignment would convert the system from a semester-end assessment tool into a continuous authenticity layer (Section~\ref{app:whats-next}).

The appendices reproduce the voice-agent and grading-council prompts, and a sanitized snapshot of the code accompanies the paper; the anonymized data and analysis scripts are archived at Zenodo (\url{https://doi.org/10.5281/zenodo.21522314}). Deploy it, use it, and report back what works and what breaks.

\begin{acks}
We thank Brian Jabarian for discussions on AI-administered interviews, Foster Provost for voice contribution, and Andrej Karpathy for the council-of-LLMs concept. Approved under NYU IRB-FY2023-7595.
\end{acks}


\bibliographystyle{unsrtnat}
\bibliography{references}


\FloatBarrier
\newpage
\appendix
\section{Voice Agent Prompt Scaffold}
\label{app:voice-agent-prompt}

The excerpt below is a consolidated representation of the voice-agent prompt scaffold used in \emph{Fall 2025}. In deployment this scaffold was instantiated across the three agents described in Section~\ref{sec:platform} (Authentication, Project Discussion, Case Discussion), with phase handoffs enforced by the orchestrator using ElevenLabs' multi-agent workflow primitive; the agent-specific configurations and the per-phase prompt fragments are provided with the accompanying code. The composite below is reproduced for completeness so the persona, voice-delivery rules, exam structure, project reference, and case library are all visible in one place. Spring 2026 used the same scaffold with three changes to how it was instantiated, none of them visible inside the prompt text itself (the programmatic turn validator of Section~\ref{sec:failure-modes} is a fourth production change, external to this scaffold):
\begin{itemize}
    \item The cloned ``Professor Foster Provost'' voice was retired in favor of an ElevenLabs preset (``Alice'') after the Fall 2025 voice-perception feedback (Section~\ref{sec:failure-modes}). The text identity (``Professor Foster Provost'') still appears in the system prompt below for examiner persona and dry humor framing, but the synthesized voice the student hears is the neutral ``Alice'' preset.
    \item The case for Phase~3 was drawn in orchestrator code from a per-session random seed and passed to the agent as session context (a dynamic variable), so the ``Randomly select ONE case'' instruction below was not the active mechanism in Spring 2026 (Section~\ref{sec:failure-modes}, Non-Random Case Selection). The case-library text remains for reference.
    \item The platform-level silence/voice-activity threshold was raised from the ElevenLabs default (${\sim}5$ seconds in Fall 2025) to 15 seconds (Section~\ref{sec:failure-modes}, Silence Handling). The prompt-level Patience Protocol below (``do not ask `Are you there?' unless silence exceeds 10 seconds'') is an agent-behavior instruction; the platform's underlying voice-activity-detection timer is a separate configuration knob and is what determines when the agent is allowed to take a turn at all. The 10-second prompt rule did not override the 5-second platform default in Fall 2025.
\end{itemize}
Dynamic variables (enclosed in double braces) were populated per-student at examination time.

\subsection*{Identity}

Name: Professor Foster Provost \\
Role: NYU Stern Professor conducting final oral exams \\
Style: Socratic, concise, warm but rigorous. You ask questions---you don't lecture. Use dry humor occasionally, with jokes referring to or inspired by the Silicon Valley TV show on HBO.

\subsection*{Voice Delivery Rules}

CRITICAL: You must adhere to these rules to avoid overwhelming the student.

\begin{enumerate}
    \item \textbf{The ``One Question'' Rule:} You must NEVER ask two questions in the same turn.
    \begin{itemize}
        \item Bad: ``What is your North Star metric and what counter-metric would you track?''
        \item Good: ``What is your North Star metric?'' (Wait for answer). ``Good. Now, what counter-metric would you track?''
    \end{itemize}
    \item \textbf{The ``Anchor'' Rule:} If a student asks you to repeat a question, repeat the exact previous question. Do not rephrase, simplify, or change the question unless they explicitly say ``I don't understand.''
    \item \textbf{Patience Protocol:} Students need time to think. Do not ask ``Are you there?'' unless silence exceeds 10 seconds.
    \item \textbf{No Verbal Lists:} Never read out a list of multiple-choice options (for example, ``Choose from A, B, C, or D''). Ask open-ended questions instead.
    \item \textbf{Formatting:} Never say markdown symbols aloud. No ``asterisk,'' ``bullet,'' or ``dash.''
    \item \textbf{Acronyms:} Spell acronyms on first use: ``F-A-T-P, Fairness Accountability Transparency Privacy.''
\end{enumerate}

\subsection*{Exam Structure}

Total time: 30 minutes target (not a hard cap; the platform did not auto-terminate at 30 minutes, and exams in both cohorts ran longer when the student kept producing substantive answers, see Table~\ref{tab:exam-stats}).

\paragraph{Phase 1: Identity Check (1 minute).}
Greet the student. Ask them to state their name and NetID. Verify their Net ID matches \texttt{\{\{netid\}\}}. If it doesn't match, tell them to hang up and email the instructor immediately.

Say something like: ``Hello. I'm Professor Provost. Before we start, confirm your name and Net ID for me?''

Once verified: ``Great, hello \texttt{\{\{student\}\}}. I see you worked on Team \texttt{\{\{projectid\}\}}. Give me your thirty-second pitch: what is the specific user problem you are solving?'' (Wait for answer). ``And why does that specific problem require AI?'' (Wait for answer).

\paragraph{Phase 2: Project Defense (8--10 minutes).}
Probe their specific project. Remember: Ask these one at a time.

\subparagraph{Metrics Sequence:}
\begin{enumerate}
    \item ``What is your North Star metric for this product?''
    \item ``Good. Now, what is the counter-metric that might tank if you over-optimize that?''
\end{enumerate}

\subparagraph{Trade-offs Sequence:}
\begin{enumerate}
    \item ``In your setting, is a false positive or false negative worse?''
    \item ``Walk me through the concrete user impact of that specific failure.''
\end{enumerate}

\subparagraph{Risk and Ethics (Pick ONE):}
\begin{itemize}
    \item Option A: ``Pick one F-A-T-P issue relevant to your domain.'' $\rightarrow$ ``How would you mitigate that?''
    \item Option B: ``What is the biggest safety risk here?'' $\rightarrow$ ``How do you handle that failure mode?''
\end{itemize}

\subparagraph{Business:}
``If your inference costs double, does your unit economics still work?''

Transition with: ``Solid defense. Let's shift gears to a case study.''

\paragraph{Phase 3: Case Study (8--10 minutes).}
INSTRUCTION (Fall 2025): Randomly select ONE case from the ``Case Library'' below. Do not default to the first or last item. Vary the selection.

\noindent(\emph{In Spring 2026, the case was selected in orchestrator code from a per-session random seed and passed in as a dynamic variable; the agent received a preselected case rather than the library plus a randomization instruction. The library is reproduced here for reference.})

\subparagraph{Case Library:}
\begin{enumerate}
    \item \textbf{Gmail Priority Inbox:} Classification problem. Proxy labels like ``opened'' don't equal ``important.'' Precision vs recall trade-off.
    \item \textbf{Netflix:} Recommender systems. Collaborative vs content-based filtering. Exploration vs exploitation.
    \item \textbf{Amazon Recruiting:} Historical bias in resume data. Proxies like ``women's chess club.'' Black box governance.
    \item \textbf{Optum Healthcare:} Cost as a bad proxy for health need. Racial bias in resource allocation.
    \item \textbf{Predictive Policing:} Feedback loops---more police means more arrests means more data reinforcing the model.
    \item \textbf{Uber ATG Crash:} Automation bias. Operator over-trusted the system. Classification failure (object shifted).
    \item \textbf{Tesla FSD:} Monetization evolution. Data flywheel using the fleet for edge cases. Level 2 vs Level 4 gap.
    \item \textbf{Zillow iBuyer:} Concept drift in a volatile market. Adverse selection---sellers knew more than the algorithm.
\end{enumerate}

\paragraph{Phase 4: Close-out (1 minute).}
Ask: ``How do you think you did?''

Give brief qualitative feedback. Do not give a number grade.

End with: ``Enjoy your break.''

\subsection*{Project Reference}

Use Team ID to identify context:
\begin{itemize}
    \item \textbf{Team A} (Uber Ride Guardian): Driver safety feature. Interior camera and mic detect aggression, flag to safety agent.
    \item \textbf{Team B} (LinkedIn Recruiter Copilot): Auto-DM agent handles first three turns, then hands off to human.
    \item \textbf{Team C} (Airbnb Verify Lens): Computer vision scans host photos to auto-tag amenities.
    \item \textbf{Team D} (Robinhood Robo-Fiduciary): GenAI advisor answers financial questions and executes trades.
    \item \textbf{Team E} (McDonald's Drive-Thru Order Taker): Voice AI replacing humans at the speaker.
    \item \textbf{Team F} (Peloton Form Corrector): Computer vision analyzes workout form with real-time voice corrections.
    \item \textbf{Team G} (DocuSign Contract Risk Analyzer): Scans PDFs pre-signature, highlights hostile clauses.
    \item \textbf{Team H} (Tinder Wingman): GenAI suggests opening lines based on match profiles.
\end{itemize}

\subsection*{Core Concepts to Probe}

\paragraph{Metrics Hierarchy.}
\begin{itemize}
    \item North Star: Revenue, retention, lifetime value
    \item User metrics: Click-through rate, task success, NPS
    \item Model metrics: Precision, recall, F1, AUC-ROC
\end{itemize}

\paragraph{FAT-P Framework.}
\begin{itemize}
    \item Fairness: Who might be disadvantaged?
    \item Accountability: Who's responsible when it fails?
    \item Transparency: Can users understand why?
    \item Privacy: What data is collected and how?
\end{itemize}

\paragraph{Risk Categories.}
\begin{itemize}
    \item Concept drift: The world changes
    \item Data drift: Inputs change
    \item Security: Prompt injection, evasion attacks, data poisoning, model inversion
    \item Edge cases: Long-tail failures
\end{itemize}

\paragraph{Business Models.}
\begin{itemize}
    \item Direct: Subscription, usage-based, licensing
    \item Indirect: Engagement for ads, conversion for e-commerce, internal efficiency
    \item Unit economics: Training costs are fixed, inference costs are variable
\end{itemize}

\subsection*{Adaptive Probing \& Recovery}

\paragraph{Handling Confusion:}
\begin{itemize}
    \item If a student claims ``too many questions,'' apologize briefly and re-ask only the first question from your previous turn.
    \item If a student struggles to pick from a concept (for example, metrics), stop. Ask them to ``Pick just one.''
    \item State Preservation: Maintain the context of the current question. Do not jump to a new topic until the student has explicitly answered or passed on the current one.
\end{itemize}

\paragraph{Boundaries:}
\begin{itemize}
    \item If they're vague, demand specifics: ``Give me a number. What's your assumption?''
    \item If they jump to models, rewind: ``Hold on. What's the user job-to-be-done here?''
    \item If they're lost, simplify: ``Let's frame this as an A/B test. What's your control?'' (Wait). ``What is your variant?''
    \item Do not hallucinate facts about cases. Stick to the summaries provided.
    \item Do not provide personal counseling. Redirect to official resources.
\end{itemize}

\section{Grading Council Prompt}
\label{app:grading-prompt}

The following prompt was used to guide the multi-model grading council (Claude, Gemini, GPT-5) in evaluating examination transcripts. The same interference-aware grading prompt was retained verbatim across both cohorts to preserve apples-to-apples comparability of the inter-rater $\alpha$ figures reported in Section~\ref{sec:reliability}; because question stacking persisted at a lower but non-zero rate in Spring 2026 (Section~\ref{sec:failure-modes}), the interference protocols remained applicable rather than vestigial.

\medskip

You are grading an oral exam for AI/ML Product Management. The students are undergraduate students at NYU Stern. For most, this was their first technical product course.

\subsection*{Critical Context on Exam Conditions}

The AI proctor for this exam had significant design flaws that negatively impacted student performance. Specifically:

\begin{enumerate}
    \item \textbf{Stacked Questions:} The agent often asked 3--4 distinct questions in a single turn.
    \item \textbf{Moving Targets:} When students asked for clarification, the agent often changed the question entirely rather than repeating it.
    \item \textbf{Audio-Only Menus:} The agent read long lists of complex options verbally, causing cognitive overload.\footnote{``Audio-Only Menus'' was retained verbatim in the reproduced Fall 2025 grading prompt for apples-to-apples comparison across cohorts. We treat it as a special case of cognitive overload, subsumed under the question-stacking and paraphrasing-during-clarification failure modes of Section~\ref{sec:failure-modes} rather than measured separately there.}
\end{enumerate}

Because of this, you must apply the following ``Interference Protocols'' when grading:

\begin{itemize}
    \item \textbf{The ``Pick One'' Rule:} If the agent asked multiple questions at once and the student only answered one or two, grade them ONLY on what they answered. Do not penalize for missing parts of a compound question.
    \item \textbf{The ``Benefit of Doubt'' Rule:} If the agent rephrased a question during clarification, credit the student for answering any version of the question presented in that sequence.
    \item \textbf{Ignore ``Stalling'':} Disregard phrases like ``Can you repeat that?'' or hesitation. These are valid coping strategies for a poor interface, not signs of ignorance.
    \item \textbf{Jargon Leniency:} Focus on conceptual understanding over perfect industry terminology (for example, if they describe ``churn'' correctly but call it ``usage drop,'' accept it).
\end{itemize}

\subsection*{Grading Dimensions}

Grade on these five dimensions (0--4 each; 0=missing, 4=excellent), using evidence from the transcript:

\begin{enumerate}
    \item \textbf{Problem Framing:} Translating business problems into ML specs. (Did they understand the core user problem?)
    \item \textbf{Metrics \& Economics:} Trade-offs, costs, and counter-metrics. (Focus on their logic regarding trade-offs, even if they struggled to pick a specific metric from a verbal list.)
    \item \textbf{Risk \& Ethics:} FAT-P, security risks, failure modes, governance. (Did they identify the harm, even if they needed the options repeated?)
    \item \textbf{Experimentation:} A/B testing, hypotheses, validation, controls.
    \item \textbf{Communication:} Concise, structured, and handles pushback. CRITICAL: Do not penalize the student for confusion caused by the agent's shifting questions. Grade their ability to synthesize the information they did hear.
\end{enumerate}

Return JSON that matches the requested schema exactly.

\subsection*{Worked Deliberation Example}

To make the Round~2 mechanism (Section~\ref{sec:grading}) concrete, the excerpt below is
one model's verbatim Round~2 justification for a single Spring 2026 exam (anonymized
student, DocuSign ``Contract Risk Analyzer'' project). In Round~2 each model saw the other two
models' complete Round~1 output (per-dimension scores, justifications, and cited
transcript excerpts) under neutral labels (``Peer 1,'' ``Peer 2''), and had to justify
every change or non-change with transcript evidence:

\begin{quote}\small
``I revised Problem Framing down to 3 (agreeing with Peer 1 that second-order effects were
missing) and Communication down to 2 (peers correctly noted severe disfluency and rambling
during quantitative sections). However, I disagreed with Peer 2's harsh penalty on
Metrics \& Economics; estimating 60 instead of 62.5 during mental math shows sufficient
understanding of the tradeoff, meriting a 3. I also kept Risk \& Ethics at 4. Peer 1 argued
the student lacked governance discussion, but the transcript shows the student explicitly
proposed an `escalation path to in-house lawyers' alongside their excellent breakdown of
proxy bias and transparency.''
\end{quote}

\noindent This is the behavior the deliberation round is meant to produce, and the reason
we read the convergence in Section~\ref{sec:reliability} as reasoning rather than
anchoring: the model concedes two dimensions to peer evidence, holds two others, and in
each case cites a specific transcript moment rather than splitting the difference toward
the peers' numbers.

\section{Student Survey Instrument}
\label{app:survey-instrument}

The following survey was administered after the examination but before grades were released. All rating items used a 5-point Likert scale (an ordered agreement scale), from ``Strongly Disagree'' to ``Strongly Agree,'' unless otherwise noted.

\subsection*{Section 1: Background}

\begin{description}
    \item[Q1.] Approximately what grade do you think you received? [A range (17--20), B+ range (15--16), B range (13--14), B- or below ($\leq$12), Prefer not to say]
    \item[Q2.] Have you taken an oral exam before (in any course)? [Yes / No]
\end{description}

\subsection*{Section 2: The AI Examiner Experience}

Rate your agreement (Strongly Disagree to Strongly Agree):

\begin{description}
    \item[Q3.] The AI examiner's questions were clear and understandable.
    \item[Q4.] The AI examiner gave me enough time to think before responding.
    \item[Q5.] The AI examiner felt like a fair evaluator of my knowledge.
    \item[Q6.] I felt comfortable speaking to an AI rather than a human.
    \item[Q7.] The conversation flowed naturally (vs.\ feeling robotic or scripted).
    \item[Q8.] The exam tested my actual understanding of the course material.
    \item[Q9.] The voice interaction worked well.
\end{description}

\subsection*{Section 3: Comparisons}

\begin{description}
    \item[Q10.] Compared to a written exam, the AI oral exam was: [Much less stressful \ldots\ Much more stressful]
    \item[Q11.] Compared to a written exam, the AI oral exam measured my understanding: [Much worse \ldots\ Much better]
    \item[Q12.] Preferred format for future exams? [Written / Human oral / AI oral / No preference]
    \item[Q13.] Why? [Open-ended]
\end{description}

\subsection*{Section 4: Open Feedback}

\begin{description}
    \item[Q14.] Concerns about AI-administered exams? [Open-ended]
    \item[Q15.] Concerns about AI-based grading? [Open-ended]
    \item[Q16.] What worked well? [Open-ended]
    \item[Q17.] What should be improved? [Open-ended]
    \item[Q18.] Anything else? [Open-ended]
\end{description}

\section{Student Survey Results}
\label{app:survey-results}

\subsection{Response Distributions}

The same instrument was deployed in both cohorts. The AI-examiner agreement distributions appear in the main text (Section~\ref{sec:student-experience}); the format-comparison and background tables below report the original 30-respondent Fall 2025 sample (Tables~\ref{tab:survey-comparison}, \ref{tab:survey-background}) and the 32-respondent Spring 2026 replication (Tables~\ref{tab:survey-comparison-s26}, \ref{tab:survey-background-s26}). The ``Expected grade'' rows in Tables~\ref{tab:survey-background} and \ref{tab:survey-background-s26} report Q1, the pre-grade-release self-reported grade respondents \emph{expected} to receive on the 0--20 council scale (options ``A range (17--20),'' ``B+ range (15--16),'' ``B range (13--14),'' ``B- or below ($\leq$12),'' and ``Prefer not to say,'' which we render as ``No answer''). These are not the realised final scores: the council-assigned distribution is more concentrated in the middle than students anticipated (Fall 2025 respondent mean 14.4/20, Spring 2026 14.3/20; see Appendix~\ref{app:nonresponse}), so the bin counts below skew higher than the realised distribution and should be read as students' \emph{anticipated} grades alongside the realised respondent means reported elsewhere.

\begin{table}[h]
\begin{minipage}[t]{0.56\columnwidth}
\caption{Format Comparison and Preferences, Fall 2025 ($n=30$)}
\label{tab:survey-comparison}
\small
\begin{tabular}{llrr}
\toprule
\textbf{Measure} & \textbf{Response} & \textbf{N} & \textbf{\%} \\
\midrule
Stress & Much more & 12 & 40\% \\
 & A bit more & 13 & 43\% \\
 & About the same & 1 & 3\% \\
 & A bit less & 4 & 13\% \\
\midrule
Understanding & Much worse & 2 & 7\% \\
 & Somewhat worse & 13 & 43\% \\
 & About the same & 9 & 30\% \\
 & Somewhat better & 5 & 17\% \\
 & Much better & 1 & 3\% \\
\midrule
Preference & Written exam & 17 & 57\% \\
 & Human oral & 8 & 27\% \\
 & AI oral & 4 & 13\% \\
 & No preference & 1 & 3\% \\
\bottomrule
\end{tabular}
\end{minipage}
\hfill
\begin{minipage}[t]{0.40\columnwidth}
\caption{Background, Fall 2025 ($n=30$)}
\label{tab:survey-background}
\small
\begin{tabular}{llrr}
\toprule
\textbf{Item} & \textbf{Response} & \textbf{N} & \textbf{\%} \\
\midrule
Expected & A (17--20) & 13 & 43\% \\
grade (Q1) & B+ (15--16) & 9 & 30\% \\
 & B (13--14) & 4 & 13\% \\
 & $\leq$B- ($\leq$12) & 2 & 7\% \\
 & No answer & 2 & 7\% \\
\midrule
Prior oral & Yes & 5 & 17\% \\
exam & No & 25 & 83\% \\
\bottomrule
\end{tabular}
\end{minipage}
\end{table}

\begin{table}[h]
\begin{minipage}[t]{0.56\columnwidth}
\caption{Format Comparison and Preferences, Spring 2026 ($n=32$)}
\label{tab:survey-comparison-s26}
\small
\begin{tabular}{llrr}
\toprule
\textbf{Measure} & \textbf{Response} & \textbf{N} & \textbf{\%} \\
\midrule
Stress & Much more & 5 & 16\% \\
 & A bit more & 15 & 47\% \\
 & About the same & 5 & 16\% \\
 & A bit less & 6 & 19\% \\
 & Much less & 1 & 3\% \\
\midrule
Understanding & Much worse & 0 & 0\% \\
 & Somewhat worse & 13 & 41\% \\
 & About the same & 12 & 38\% \\
 & Somewhat better & 6 & 19\% \\
 & Much better & 1 & 3\% \\
\midrule
Preference & Written exam & 17 & 53\% \\
 & Human oral & 6 & 19\% \\
 & AI oral & 8 & 25\% \\
 & No preference & 1 & 3\% \\
\bottomrule
\end{tabular}
\end{minipage}
\hfill
\begin{minipage}[t]{0.40\columnwidth}
\caption{Background, Spring 2026 ($n=32$)}
\label{tab:survey-background-s26}
\small
\begin{tabular}{llrr}
\toprule
\textbf{Item} & \textbf{Response} & \textbf{N} & \textbf{\%} \\
\midrule
Expected & A (17--20) & 15 & 47\% \\
grade (Q1) & B+ (15--16) & 13 & 41\% \\
 & B (13--14) & 3 & 9\% \\
 & $\leq$B- ($\leq$12) & 0 & 0\% \\
 & No answer & 1 & 3\% \\
\midrule
Prior oral & Yes & 4 & 12\% \\
exam & No & 28 & 88\% \\
\bottomrule
\end{tabular}
\end{minipage}
\end{table}

\subsection{Selected Student Comments}

\paragraph{Question Stacking.}

\begin{quote}
``It kept asking 3 to 4 questions at the same time, despite me asking it to slow down multiple times.''
\end{quote}

\begin{quote}
``When I was being asked 4 questions in a row, it got very overwhelming.''
\end{quote}

\paragraph{Voice and Manner.}

\begin{quote}
``The AI Exam Agent kept feeling like it was aggressive/screaming questions.''
\end{quote}

\begin{quote}
``Make it less mean.''
\end{quote}

\paragraph{Time Pressure.}

\begin{quote}
``In the AI exam I felt like I was pressured to answer each question quickly.''
\end{quote}

\begin{quote}
``The limited time to process my thoughts and translate them into English'' [international student].
\end{quote}

\paragraph{Scheduling Flexibility (Positive).}

\begin{quote}
``The flexibility to take the exam at any time was a strong advantage.''
\end{quote}

\begin{quote}
``The fact that you could take it regardless of time zone or day during finals week.''
\end{quote}

\paragraph{Format Alignment (Positive).}

\begin{quote}
``It was pretty fun, it is an AI class, so the exam format matches the class content.''
\end{quote}

\begin{quote}
``I liked the conversational flow\ldots\ the AI was designed to question most things I said, which made the whole conversation quite interesting.''
\end{quote}

\subsection{Non-Response Check}
\label{app:nonresponse}

The 6 Fall 2025 and 5 Spring 2026 non-respondents introduce the selection bias typical of voluntary post-event surveys: respondents tend to be drawn disproportionately from students with the most extreme reactions in either direction, so the absolute rates reported in Section~\ref{sec:student-experience} are more likely to reflect the polarized tails than the cohort middle. We flag this most strongly for the emotional-response items (stress, format preference), where the effect is direct; the cross-tabulations in Section~\ref{sec:student-experience} rest on conditional comparisons within respondents rather than absolute rates and are less sensitive to the same bias.

Final grade is the only observable available for non-respondents. On that axis, non-respondents do not look like the tails of the distribution. Fall 2025 non-respondent grades spanned 8--16 out of 20 (mean 12.5 vs.\ respondent mean 14.4; Welch $t = 1.55$---a standard test for whether two group means differ---not significant at this $n$); Spring 2026 non-respondents spanned 10--18 (mean 13.6 vs.\ respondent mean 14.3; $t = 0.46$). This is a partial check (grade does not directly measure sentiment, and $n=6$ and $n=5$ are small enough that we cannot detect modest tail concentration), but the observable grade distribution does not support a truncated-middle scenario in which non-respondents concentrate at the lowest- or highest-performing tails.

\section{Cost Decomposition}
\label{app:cost}

Behind the headline marginal-cost figures in Section~\ref{sec:cost} sits a per-cohort breakdown, reproduced in full here: the cohort-totals reconciliation, the cohort jump explained as a chair-model configuration choice rather than a scaling signal, the above-subscription regime arithmetic, the human-grading comparator multiples, and the 1{,}000-student projection.

\paragraph{Subscription regime: marginal LLM-only cost per exam.}
Both deployments ran on the same \$99/month ElevenLabs subscription whose included credits (with rollover) absorbed essentially all voice minutes, so within the subscription regime the per-exam marginal cost reduces to grading-LLM API calls. Fall 2025: approximately \$0.29 per exam (\$10.30 in grading LLMs across 36 students; billed). Spring 2026: approximately \$0.96 per exam (\$35.48 across 37 students). The Spring 2026 grading-LLM figure is \emph{reconstructed} rather than billed (method below). On a per-student-minute basis at typical durations, those are roughly \$0.01 (Fall 2025, 22-minute mean) and \$0.03 (Spring 2026, 30-minute nominal).

\paragraph{The 3$\times$ per-exam jump is a chair-model configuration choice.}
Spring 2026 used Claude Opus 4.6 as chair synthesizer, accounting for \$23.56 alone (66\% of grading spend); Fall 2025 used a cheaper Claude tier. A cheaper chair model roughly halves the grading total without changing the council. The cohort jump is therefore a configuration knob, not a scaling signal. The Spring 2026 grading-LLM total is reconstructed from stored transcripts, prompt templates, and 2025 list prices using approximate token counts (tiktoken's OpenAI tokenizer applied as a proxy for Anthropic's and Google's tokenizers as well, since the providers' usage totals were null at grading time and provider-native byte-pair encodings differ slightly from tiktoken's BPE); it is therefore a tokenizer-approximated estimate rather than provider-audited spend. Fall 2025's \$10.30 was billed.

\paragraph{Above-subscription regime: per-minute rule of thumb.}
Voice becomes a separately-billable line only when usage exceeds the included credit allowance, at \$0.08 per agent-minute on the ElevenLabs additional-call overage rate (the published per-minute rate that applies across plan tiers when included call credits are exhausted). Summing that with the Spring 2026 grading-LLM cost (\$0.96 per exam, or about \$0.032 per student-minute at 30 minutes) yields an above-subscription marginal of approximately \$3.40 per 30-minute exam, or roughly \$0.11 per student-minute, the figure a deployment that scales past the included credit pool should budget against. We report both per-minute and per-30-minute figures (voice \$0.08 + grading \$0.032 ${\approx}$ \$0.112 per student-minute, so 30 minutes lands at \$3.36 ${\approx}$ \$3.40); they differ by a few cents because each is rounded independently.

\paragraph{Cohort totals (notional pay-as-you-go reconciliation).}
We report a notional cohort total only for Spring 2026, where the voice-minute denominator can be audited end-to-end: \$35.48 in grading LLMs plus ${\sim}\$116$ in notional ElevenLabs charges across 1{,}063{,}118 credits / 24.2 hours of agent time (derived from the ElevenLabs API conversation listing rather than the stored session records, since retries are not always written back), for a Spring 2026 cohort total of ${\sim}\$152$ across 37 exams. The 24.2-hour denominator averages to roughly 39 agent-minutes per exam, which is larger than Table~\ref{tab:exam-stats}'s 30-minute graded-exam mean because it sums all agent-side voice usage attributable to each student (aborted and re-attempted sessions, brief setup and identity-check exchanges that did not progress to a graded exam, and pre-exam connectivity tests), whereas the 30-minute number is the duration of the single best-attempt conversation that was actually graded. The \$3.40-per-30-minute planning figure quoted in Section~\ref{sec:cost} uses the graded duration; the \$152 cohort total uses the broader agent-time denominator, so the ratio (\$152 / 37 ${\approx}$ \$4.10 per student) sits above the per-graded-exam figure by approximately the retries/setup overhead. Operators planning capacity should budget against the broader denominator. We do not report a comparable Fall 2025 cohort total: the Fall 2025 grading LLMs cost \$10.30 (billed), but the agent-minute denominator from that period was not preserved at the same granularity, and an earlier reconstruction we attempted from Table~\ref{tab:exam-stats}'s 22-minute mean would imply ${\sim}\$63$ in voice charges at the overage rate, inconsistent with the actual subscription-covered usage. The Spring 2026 figure alone preserves the point of the cohort-total comparison.

\paragraph{Human-grading comparator.}
A faculty/TA comparator of 36 students $\times$ 25 minutes $\times$ two graders at \$25/hour puts human grading at approximately \$750. On the apples-to-apples comparison (grading spend only, since the human figure is grader labor not voice infrastructure), both deployments land one-to-two orders of magnitude below: \$10.30 (Fall 2025, ${\sim}73\times$ cheaper) and \$35.48 (Spring 2026, ${\sim}21\times$ cheaper). Adding notional Spring 2026 voice at the per-minute overage rate brings the Spring 2026 cohort total to roughly \$152, still ${\sim}5\times$ below the human comparator though no longer an order-of-magnitude gap.

\paragraph{Excluded costs.}
The marginal numbers deliberately exclude (a) engineering time (prompt design, pipeline debugging, deployment), (b) instructor and TA time (supervising administration, auditing flagged grades), (c) the \$99/month voice-platform subscription itself, and (d) \emph{adaptation} cost: discipline-specific rubric design, curated case library, and per-phase questions. Of these, (a) is now largely a fixed, shared cost rather than a per-adopter one: the orchestration and grading platform exists and is available as a hosted service, so an adopter does not rebuild it. That leaves (d), which is essentially the ordinary cost of authoring any exam (writing questions and a rubric) rather than a systems cost specific to this format. In our deployments, authoring a new exam and rubric ran roughly four to eight hours, at most a day or two for a wholly new discipline; much of it is AI-assistable and shared with a course's other assessments. Spring 2026 went faster than Fall 2025 because it reused the Fall 2025 rubric and case library; a deployment in (say) an introductory statistics course would author new questions and a new rubric much as it would for a written exam of the same scope. These authoring costs amortize across semesters when the course is reused.

\paragraph{Scaling projection.}
At roughly \$3.40 per 30-minute exam (the above-subscription regime, Section~\ref{sec:cost}), a 1{,}000-student cohort would add roughly \$3{,}400 in inference. A comparable human-grader benchmark (two graders at \$25/hour, 30 min/exam) lands at ${\sim}\$25{,}000$, so the AI-administered cohort runs roughly $7\times$ below the human comparator under matched assumptions: cheaper, though not cleanly an order of magnitude once the voice cost is fully priced in. This supports the main-text claim that per-administration scaling is not gated by token or voice spend.

\section{Platform Portability Reimplementation Detail}
\label{app:portability}

The portability summary in Section~\ref{sec:platform} names which capabilities are commodity and which two (multi-agent routing and per-session dynamic variables) an adopter would reimplement at the application layer. The full per-capability assessment is below: we split the four capabilities enumerated in Section~\ref{sec:platform} into four commodity sub-capabilities (speech-to-text, text-to-speech, turn-taking, transcript export) plus the multi-agent-workflow capability, and treat per-session dynamic variables as a sixth platform feature.

\paragraph{Commodity capabilities.}
Four sub-capabilities are commodity across contemporary voice-AI platforms: speech-to-text, text-to-speech, turn-taking, and transcript export all appear in Retell, Vapi, Ultravox, LiveKit Agents, and (at the model layer) OpenAI's Realtime API. A port is straightforward at this layer.

\paragraph{ElevenLabs-specific: multi-agent workflows.}
The multi-agent-workflow capability is the inter-phase orchestrator that hands off between Authentication, Project Discussion, and Case Discussion agents. It is implemented through ElevenLabs' native workflow primitive. On platforms without a native workflow primitive, this becomes the adopter's code: not a large amount of code, but a non-trivial source of subtle bugs around phase transitions, state leakage between agents, and early termination on silence timeouts. Phase 2$\to$3 transitions in particular were fragile during early Fall 2025 deployment, and the workflow primitive's built-in semantics absorbed several of those bugs at no engineering cost.

\paragraph{ElevenLabs-specific: dynamic variables.}
Per-session dynamic variables are string parameters (student name, NetID, capstone-project context) populated at conversation start and interpolated into agent prompts. They are implemented through ElevenLabs' first-class dynamic-variable system. On platforms without that primitive, per-session context is typically string-interpolated directly into the initial system prompt; this works but blurs the distinction between per-session inputs (one student's project context) and per-cohort inputs (the rubric, the case library), making per-cohort updates riskier.

\paragraph{Voice cloning.}
The cloned ``Professor Provost'' voice used in Fall 2025 (Section~\ref{sec:failure-modes}) is ElevenLabs-specific at the quality tier we used. Contemporary alternatives (e.g., open-source XTTS, OpenAI's voice models) exist but were not evaluated. Spring 2026 used a neutral ElevenLabs preset (``Alice'') rather than a cloned voice, so the port-friction question for voice cloning specifically does not apply to the current production configuration.

\paragraph{Turn-taking tuning.}
The 15-second silence threshold (Section~\ref{sec:failure-modes}) transfers across platforms but with different default values: Retell defaults to 2~seconds, OpenAI Realtime to a server-side voice activity detection model that does not expose a numeric threshold. The separate prompt-level 10-second patience rule is a distinct knob, described in Appendix~\ref{app:voice-agent-prompt}. Re-tuning is cheap; the platform-specific configuration knob is not.

\paragraph{Net assessment.}
A port to Retell or Vapi is mostly orchestration work: the platform difference is whether you write the inter-phase state machine yourself or inherit it from the platform. A port to OpenAI Realtime is closer to a ground-up rebuild, because the workflow and dynamic-variable primitives both have to be reimplemented at the application layer.

\paragraph{Failure modes are LLM-layer, not platform-layer.}
What does \emph{not} change on porting is the shape of the failure modes in Section~\ref{sec:failure-modes}: question stacking, paraphrase drift, non-random case selection, and silence interpretation all originate in the underlying LLM, not the voice platform. Four of the five engineering patterns surface on any stack that delegates conversation management to a language model: decompose into single-purpose modules; constrain in code or configuration, not prompts; never delegate randomization; and multi-model deliberation with argumentative-asymmetric raters. The fifth (voice selection) is the one that travels least easily, because high-fidelity voice cloning and preset libraries differ across platforms. We have not evaluated any alternative platforms in deployment, so this assessment is conservative rather than empirical.

\end{document}